\documentclass[%
 reprint,
superscriptaddress,
 amsmath,amssymb,
 aps,
]{revtex4-2}

\usepackage{graphicx}
\usepackage{dcolumn}
\usepackage{bm}
\usepackage{tikz}
\usetikzlibrary{decorations.pathreplacing,decorations.pathmorphing}
\usepackage{hyperref}

\begin{document}

\title{Generating many-component cat states and their Bell pairs}

\author{Ari John Boon}
\email{ari-john.boon@polymtl.ca}
\affiliation{Department of Engineering Physics, École polytechnique de Montréal, Montréal, QC, H3T 1J4, Canada}

\author{Olivier Landon-Cardinal}
\affiliation{École de technologie supérieure, Université du Québec, Montréal, QC, H3C 1K3, Canada}

\author{Nicol{\'a}s Quesada}
\affiliation{Department of Engineering Physics, École polytechnique de Montréal, Montréal, QC, H3T 1J4, Canada}

\date{\today}

\begin{abstract}

To counter loss errors in a quantum photonic system, we had previously found substantial performance advantages in using many-component cat states compared to their common two-component counterparts,
and here we propose both probabilistic and deterministic protocols for the generation of these states.
Our probabilistic schemes are demonstrated analytically, with a non-zero lower bound on the probability of generating perfect single-mode cat states in the limit of large amplitude, using the Kerr gate for the production of resource states.
Our deterministic schemes rely on the universality of Kerr and displacement gates, with characteristics that appear uniquely suited to the generation of many-component cat states.
For cat-encoded Bell states, we find they can be generated via a generalisation of the Kerr and displacement method, or via post-selection from two combined many-component cat states.

\end{abstract}

\maketitle

\section{Introduction}

Loss is an inherent aspect of photonic systems, and will destroy quantum information in the absence of quantum error correction.
Various bosonic codes have been developed or found to be tolerant to loss \cite{gottesman2000encoding,grimsmo2020quantum,bergmann2016quantum,albert2018performance}, and now there is growing interest in finding states that not only belong to loss-tolerant codes, but are also more practical to generate or operate \cite{fur2026deterministic, hanks2026use, wetherbee2026bosonic}.
No matter the theoretical corrective capacity of a code, its true efficacy is determined in large part by how well it can be implemented, \textit{e.g.}, how well the necessary states can be produced.
It is by taking advantage of the inherent aspects of bosonic systems that practical gains can be made in both the performance and simulation of these systems \cite{upreti2026exponentially, guseynov2026coherent, martinez2026calculating, marshall2023simulation}.
The choice of the coherent state basis in particular is attractive, given that the accounting for Gaussian operations and photon-number measurements is relatively easy, while the Kerr nonlinearity has a convenient decomposition into Gaussian gates \cite{tara1993production,wu1996superposition}.
While nonlinearities can be challenging or resource-intensive to implement, particularly in optics \cite{shapiro2006single}, the Kerr gate's decomposition shows interesting symmetry characteristics, similar to those seen in cat states and other rotationally-symmetric bosonic (RSB) states \cite{grimsmo2020quantum}.
Beyond their capacity for loss tolerance, RSB states---including cat states---have also been of interest as resources for the generation of Gottesman-Kitaev-Preskill (GKP) states \cite{weigand2018generating, zheng2023gaussian, kiryu2026linear}.

Prior proposals for the generation of cat states in particular have largely been focused on their lowest-order forms, specifically superpositions of two or four coherent states \cite{glancy2008methods, thekkadath2020engineering, takase2021generation, zhang2022all, noury2025simple}, which have been successfully produced in experiment \cite{ourjoumtsev2007generation,chen2024generation}.
The higher-order counterparts to these cats have been relatively unexplored, in spite of the raw performance benefits that can be achieved with them \cite{albert2018performance, boon2026generalised}.
Where the generation of superpositions of many coherent states has been considered, often utilising the Kerr gate \cite{tara1993production, adam2011double}, the states produced have not necessarily had the rotational symmetry essential to the structure of cat states \cite{lin2026preparing}, and therefore lack characteristics such as the periodic Fock support that grants cat states loss tolerance \cite{bergmann2016quantum}.
As the capacity to implement tunable optical Kerr gates improves \cite{costanzo2017measurement}, it becomes an important tool to consider in the generation of useful states, particularly with respect to cats.

Here, we show that Kerr gates can produce resource states for interferometric cat state generation, allowing for the generation of cat states with arbitrary component numbers beyond the two or four typically seen in the literature.
We further generalise the protocol of Su et al. \cite{su2022universal} to convert those states into cat-encoded Bell states.
Because of the cat code's native definition in the coherent state basis, and the decomposition of the Kerr gate into rotations, this can all be demonstrated analytically.

We have also shown how deterministic generation via stages of displacement and Kerr gates can produce high-quality many-component cat states.
The combination of displacement and Kerr gates is a universal approach to state generation, but we have found features that suggest that it may be uniquely well-suited to the generation of many-component cat states, and which reduce the numerical overhead in optimising the gates.
Similarly, we propose a two-mode scheme with the introduction of beamsplitters that appears promising as a source of deterministic cat-encoded Bell states.

The outline of our paper is as follows.
In Sec.~\ref{sec:cats} we define the cat states and elaborate on operations in the coherent state basis.
In Sec.~\ref{sec:kerr} we introduce the Kerr gate, including its decomposition into rotations.
In Sec.~\ref{sec:prob} we outline our two interferometric protocols: in Sec.~\ref{sec:KerrDFT} for $q$-component cat states via Kerr gates and a DFT interferometer, and in Sec.~\ref{sec:bell} for cat-encoded Bell states produced via interferometry on two single-mode cat states.
In Sec.~\ref{sec:deterministic} we show high-quality single-mode $q$-component cat state generation via a combination of iterated displacement and Kerr gates, and an extension of the scheme for Bell cats via the addition of beamsplitters.
In Sec.~\ref{sec:conclusion} we provide our outlook on the results we have shown, in particular on the prospects for improved techniques surrounding state generation.

\section{Multicomponent cat states}
\label{sec:cats}

One may define a logical basis from rings of even numbers of coherent states, $\vert \alpha \rangle$,
\begin{align}
\big\vert \tilde{0}_L \big\rangle
&=
\frac{1}{\sqrt{\mathcal{N}_0}}
\sum^{2 L + 1}_{m=0}
\vert e^{i m \phi_L} \alpha \rangle,
\label{eqn:codewords0}
\\
\big\vert \tilde{1}_L \big\rangle
&=
\frac{1}{\sqrt{\mathcal{N}_1}}
\sum^{2 L + 1}_{m=0}
(-1)^m
\vert e^{i m \phi_L} \alpha \rangle,
\label{eqn:codewords1}
\end{align}
where $\phi_L \equiv \pi / (L+1)$ and ${\mathcal{N}_{0}}$, ${\mathcal{N}_1}$ are normalisation constants.
This follows the definition of RSB states given by Grimsmo et al. \cite{grimsmo2020quantum}, where the rotation $\hat{R}(\phi_L)$ acts as a logical Pauli $Z$ gate.
These states exhibit a combed structure in Fock space, giving them a tolerance $L$ to discrete loss events \cite{bergmann2016quantum}, which is a property of RSB codes in general \cite{grimsmo2020quantum}.
More precisely, cat states of order $L$ can recover from the first $L$ Krauss operators of the loss channel.

Coherent states themselves are defined as
\begin{equation}
\vert \alpha \rangle
=
e^{-|\alpha|^2/2}
\sum^{\infty}_{n=0}
\frac{\alpha^n}{\sqrt{n!}}
\vert n \rangle,
\label{eqn:coherent}
\end{equation}
where here $\alpha \in \mathbb{C}$, and $\vert n \rangle$ is a Fock state of photon-number $n$.
Coherent states are eigenstates of the annihilation operator,
\begin{equation}
\hat{a} \vert \alpha \rangle
=
\alpha \vert \alpha \rangle,
\label{eqn:aa}
\end{equation}
and have a mean photon-number of
\begin{equation}
\langle \alpha \vert \hat{n} \vert \alpha \rangle = \vert \alpha \vert^2,
\end{equation}
where $\hat{n} = \hat{a}^{\dagger} \hat{a}$.
Moreover, coherent states are a convenient basis to work in because some common operations transform them succinctly, such as displacement,
\begin{align}
\hat{D}(\beta) \vert \alpha \rangle
&=
\exp{(\beta \hat{a}^{\dagger} - \beta^* \hat{a})} \vert \alpha \rangle
\\
&= 
\exp{\left[(\beta \alpha^* - \beta^* \alpha)/2\right]}
\vert \alpha + \beta \rangle,
\label{eqn:displacement}
\end{align}
and rotation,
\begin{equation}
\hat{R}(\phi) \vert \alpha \rangle
= e^{i \phi \hat{n}} \vert \alpha \rangle
= \vert e^{i\phi} \alpha \rangle,
\end{equation}
while photon-number measurement can also be described succinctly and offers a source of nonlinearity,
\begin{equation}
\langle n \vert \alpha \rangle
=
e^{-|\alpha|^2/2} \frac{\alpha^n}{\sqrt{n!}}.
\end{equation}
Similarly, the overlap of two coherent states is described simply by \cite{gerry2023introductory}
\begin{equation}
\langle \alpha \vert \gamma \rangle
=
{\exp}{\left[ \frac{1}{2} \left( \alpha^* \gamma - \alpha \gamma^* - \vert \alpha - \gamma \vert^2 \right) \right]},
\label{eqn:coherent_overlap}
\end{equation}
while more generally we will define the fidelity of two states as
\begin{equation}
F = \vert \langle \psi \vert \phi \rangle \vert^2.
\label{eqn:fidelity}
\end{equation}
We will at times define interferometers within mode space but use their Hilbert space representation in our expressions; the latter representation will be denoted via a hat, such that if we have a passive unitary interferometer parametrised by matrix $V$ in mode space, then the associated Hilbert space operator will be denoted as $\hat{V}$, given by
\begin{equation}
\hat{V} = \hat{\mathcal{U}}(V),
\end{equation}
where for $n$ modes \cite{miatto2020fast}
\begin{align}
\hat{V}^{\dagger} a^{\dagger}_i \hat{V}
=
\sum^n_{j=0} V^*_{ij} a^{\dagger}_j,
\quad
\hat{V} a^{\dagger}_j \hat{V}^{\dagger}
=
\sum^n_{i=0} V^*_{ij} a^{\dagger}_i,
\end{align}
which can be tracked in the coherent state basis easily via $V$ alone.
For example, given the two-mode parametrisation
\begin{equation}
V
=
\begin{pmatrix}
V_{00} & V_{01} \\
V_{10} & V_{11}
\end{pmatrix},
\end{equation}
and coherent states $\vert \alpha, \beta \rangle$, the transformed state can be found via
\begin{equation}
V
\begin{pmatrix}
\alpha \\
\beta
\end{pmatrix}
=
\begin{pmatrix}
V_{00} \alpha + V_{01} \beta \\
V_{10} \alpha + V_{11} \beta
\end{pmatrix},
\end{equation}
meaning
\begin{equation}
\hat{V} \vert \alpha, \beta \rangle
=
\vert
V_{00} \alpha + V_{01} \beta,
V_{10} \alpha + V_{11} \beta
\rangle.
\end{equation}
The numerical convenience of the coherent state basis has received growing attention \cite{dias2024classical, solodovnikova2025fast, upreti2026exponentially, guseynov2026coherent}, and because cat states are natively defined within it they make attractive candidates for study.

We will also consider cat-encoded Bell states, albeit of a form with a deformation, \textit{e.g.},
\begin{equation}
\vert \tilde{\Phi}^{+}_L \rangle
=
\frac{1}{\sqrt{2}}
\left(
\vert \tilde{0}_L, \tilde{0}_L \rangle
+
\frac{\mathcal{N}_1}{\mathcal{N}_0}
\vert \tilde{1}_L, \tilde{1}_L \rangle
\right),
\label{eqn:bell_deformed}
\end{equation}
where for $\alpha \rightarrow \infty$, the ratio of the normalisation constants goes to one, \textit{i.e.}, $\mathcal{N}_1 / \mathcal{N}_0 \rightarrow 1$.
The mis-match of the normalisation constants stems from choosing the logical basis defined by Eqs. \eqref{eqn:codewords0} and \eqref{eqn:codewords1}, and in Appendix~\ref{app:bell} we show how the deformation appears in the Bell state.
This choice of basis enforces orthogonality for any value of $\alpha$ but introduces deformation by way of the codeword normalisation mis-match, which we discussed in our previous work \cite{boon2026generalised}:
during error correction, the normalisation constants in the codewords are re-distributed within the logical space of the code, and as such if they are not equal they deform the transmitted qubit.
At $L=1$ the Bell state we have defined is identical to that used by Refs.~\cite{bergmann2016quantum,hastrup2022all,su2022universal}, though when written in their basis does not exhibit the deformation, instead trading it for non-orthogonality of the codewords.
However, in the limit of $|\alpha| \gg 1$, these two choices of basis become approximately equal.

Note also that we will refer sometimes to $q$-component cat states more generally, which encompasses both Eqs.~(\ref{eqn:codewords0}) and (\ref{eqn:codewords1}), but also states with odd numbers of components, for which this logical basis is not defined.
In certain cases then we will use $q$-component cat states defined as
\begin{equation}
\vert \text{cat}_q (\alpha) \rangle
=
\frac{1}{\sqrt{\mathcal{N}}}
\sum^{q-1}_{m=0}
\vert e^{2 \pi i / q} \alpha \rangle,
\end{equation}
which are equal to $\vert \tilde{0}_L \rangle$ for $q=2(L+1)$.

\section{The Kerr gate}
\label{sec:kerr}

All of the techniques we describe will utilise the Kerr gate, which we will define as
\begin{equation}
\hat{K}(\kappa)
=
\exp{\left( i \pi \kappa \hat{n}^2 \right)}.
\label{eqn:kerr}
\end{equation}
The first interesting property of the Kerr gate that we will use is its decomposition for rational $\kappa$ into a linear combination of rotations \cite{tara1993production, wu1996superposition, upreti2026exponentially},
\begin{equation}
\hat{K}\left(-\frac{p}{q}\right)
=
\sum^{q-1}_{j=0}
g_{j,p,q}
\,
\exp{\left( -i \frac{2 \pi j}{q} \hat{n} \right)},
\label{eqn:K_decomp}
\end{equation}
where
\begin{equation}
g_{j,p,q}
=
\frac{1}{q}
\sum^{q-1}_{k=0}
\exp{
\left(
\frac{\pi i k}{q}
\left[
2j
-
pk
+
\frac{(-1)^{pq}-1}{2}
\right]
\right)},
\label{eqn:gj}
\end{equation}
which we show in Appendix~\ref{app:sumgauss} can be written explicitly, for $p=\pm1$, as
\begin{equation}
g_{j,\pm 1,q}
=
\frac{1}{\sqrt{q}}
{\exp}{\left[
\pm
\frac{i \pi}{q}
\left(
j - \frac{(-1)^q - 1}{4}
\right)^2
\right]}
{\exp}{\left[
\mp
\frac{i \pi}{4}
\right]}.
\label{eqn:gj_s}
\end{equation}
This decomposition is essential to both of our proposals for the generation of $q$-component cat states in Secs.~\ref{sec:KerrDFT} and \ref{sec:deterministic}.
Notably, Kerr gates $\hat{K}(-1/2)$ can function as phase gates in the cat codespace spanned by Eqs.~(\ref{eqn:codewords0}) and (\ref{eqn:codewords1}) for $L$ even \cite{grimsmo2020quantum}, and acts trivially on $L$ odd, where $2(L+1)=q$;
we know
\begin{equation}
\hat{K}(-1/2)
=
\frac{(1+i)\hat{R}(0) + (1-i)\hat{R}(\pi)}{2},
\end{equation}
and
\begin{align}
(1+i) + (1-i) &= 2, \\
(1+i) - (1-i) &= 2i,
\end{align}
so for $L$ even, terms in $\vert \tilde{0}_L \rangle$ have no sign difference and are unchanged by $\hat{K}(-1/2)$, while for $\vert \tilde{1}_L \rangle$ any terms related by $\hat{R}(\pi)$ are different by a sign and accrue an overall factor $i$;
for $L$ odd, all terms in either codeword have no sign difference from their $\hat{R}(\pi)$ counterpart, and thus $\hat{K}(-1/2)$ behaves identically on both codewords.

Kerr gates applied to coherent states, as
\begin{equation}
\hat{K}\left(-\frac{p}{q}\right) \vert \alpha \rangle
=
\sum^{q-1}_{j=0}
g_{j,p,q}
\vert
e^{-2 i \pi j / q}
\alpha
\rangle,
\end{equation}
are known as Kerr states, and are superficially similar to $q$-component cat states, having equivalent rotations.
However because of their coefficients $g_{j,p,q}$ they do not demonstrate the necessary rotational symmetry and thus do not have the comb-like Fock structure that gives cat states their loss tolerance.

More generally, if one has access to Kerr gates along with Gaussian gates, one has a universal set of operations for continuous-variable quantum computation \cite{lloyd1999quantum}.
In particular, Kerr and displacement gates are sufficient for universality \cite{guseynov2026coherent}, and will be the gates we use for the deterministic generation of cat states in Sec.~\ref{sec:deterministic}.

\section{Probabilistic generation schemes}
\label{sec:prob}

Here we will demonstrate generation schemes based on interferometry and post-selection, where for single-mode cat states we use Kerr states as a resource, while Bell pairs we produce from two cat state resources.
While both of these methods are inherently probabilistic, they benefit from being demonstrably perfect generation schemes under certain conditions: for single-mode cat states in the limit of large amplitude, and for Bell pairs when the resource cat states are themselves perfect.
In the former case, while approaching the limit of large amplitudes is ultimately infeasible, we find that fidelity improves monotonically with amplitude, and thus high fidelities may still be achieved at finite amplitudes.
We also demonstrate the behaviour and implementation of the Fock damping operator \cite{tzitrin2020progress} in the coherent state basis, which is necessary for the correct operation of the telecorrection schemes we are interested in, though it comes at a cost both in further operations and in probability.
With these three techniques it is possible to generate all of the requisite states to perform loss correction on a many-component cat qubit \cite{hastrup2022all, boon2026generalised}.

\subsection{$q$-cats via Kerr-DFT}
\label{sec:KerrDFT}

We introduce our scheme for probabilistically generating states whose fidelity to cat states improves with the amplitude, $|\alpha|$, becoming perfect-fidelity $q$-component cat states in the limit of large amplitude, while the success probability approaches a non-zero lower bound.
Our scheme uses the following elements:
\begin{itemize}
\item $q$ coherent states, $\vert \beta \rangle$, where $\beta = \alpha/ \sqrt{q}$,
\item $q$ Kerr gates, $\hat{K}(-1/q)$,
\item a $q$-mode DFT interferometer,
\item $q-1$ threshold detectors.
\end{itemize}
It is worth noting that the DFT interferometer has a low-depth linear-optical implementation \cite{barak2007quantum}, and that our protocol requires only threshold detection, which is another boon.

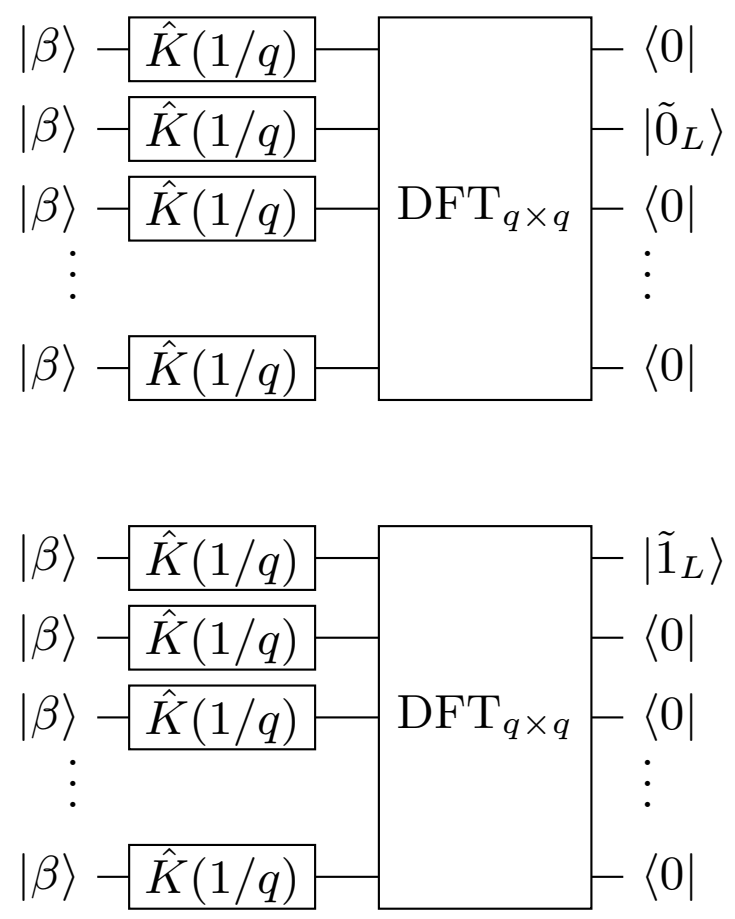
\begin{figure}[t]
\begin{tikzpicture}[scale=1.000000,x=1pt,y=1pt]
\filldraw[color=white] (0.000000, -7.500000) rectangle (99.000000, 67.500000);
% Drawing wires
% Line 1: a W {\vert \beta \rangle} {\langle 0 \vert}
\draw[color=black] (0.000000,60.000000) -- (99.000000,60.000000);
\draw[color=black] (0.000000,60.000000) node[left] {${\vert \beta \rangle}$};
% Line 3: b W {\vert \beta \rangle} {\vert \tilde{0}_L \rangle}
\draw[color=black] (0.000000,45.000000) -- (99.000000,45.000000);
\draw[color=black] (0.000000,45.000000) node[left] {${\vert \beta \rangle}$};
% Line 5: c W {\vert \beta \rangle} {\langle 0 \vert}
\draw[color=black] (0.000000,30.000000) -- (99.000000,30.000000);
\draw[color=black] (0.000000,30.000000) node[left] {${\vert \beta \rangle}$};
% Line 7: ...d W
\draw[color=black] (0.000000,15.000000) node[anchor=mid east] {$\vdots$};
% Line 8: e W {\vert \beta \rangle} {\langle 0 \vert}
\draw[color=black] (0.000000,0.000000) -- (99.000000,0.000000);
\draw[color=black] (0.000000,0.000000) node[left] {${\vert \beta \rangle}$};
% Done with wires; drawing gates
% Line 2: a G $\hat{K}(1/q)$ width=35
\begin{scope}
\draw[fill=white] (23.500000, 60.000000) +(-45.000000:24.748737pt and 8.485281pt) -- +(45.000000:24.748737pt and 8.485281pt) -- +(135.000000:24.748737pt and 8.485281pt) -- +(225.000000:24.748737pt and 8.485281pt) -- cycle;
\clip (23.500000, 60.000000) +(-45.000000:24.748737pt and 8.485281pt) -- +(45.000000:24.748737pt and 8.485281pt) -- +(135.000000:24.748737pt and 8.485281pt) -- +(225.000000:24.748737pt and 8.485281pt) -- cycle;
\draw (23.500000, 60.000000) node {$\hat{K}(1/q)$};
\end{scope}
% Line 4: b G $\hat{K}(1/q)$ width=35
\begin{scope}
\draw[fill=white] (23.500000, 45.000000) +(-45.000000:24.748737pt and 8.485281pt) -- +(45.000000:24.748737pt and 8.485281pt) -- +(135.000000:24.748737pt and 8.485281pt) -- +(225.000000:24.748737pt and 8.485281pt) -- cycle;
\clip (23.500000, 45.000000) +(-45.000000:24.748737pt and 8.485281pt) -- +(45.000000:24.748737pt and 8.485281pt) -- +(135.000000:24.748737pt and 8.485281pt) -- +(225.000000:24.748737pt and 8.485281pt) -- cycle;
\draw (23.500000, 45.000000) node {$\hat{K}(1/q)$};
\end{scope}
% Line 6: c G $\hat{K}(1/q)$ width=35
\begin{scope}
\draw[fill=white] (23.500000, 30.000000) +(-45.000000:24.748737pt and 8.485281pt) -- +(45.000000:24.748737pt and 8.485281pt) -- +(135.000000:24.748737pt and 8.485281pt) -- +(225.000000:24.748737pt and 8.485281pt) -- cycle;
\clip (23.500000, 30.000000) +(-45.000000:24.748737pt and 8.485281pt) -- +(45.000000:24.748737pt and 8.485281pt) -- +(135.000000:24.748737pt and 8.485281pt) -- +(225.000000:24.748737pt and 8.485281pt) -- cycle;
\draw (23.500000, 30.000000) node {$\hat{K}(1/q)$};
\end{scope}
% Line 9: e G $\hat{K}(1/q)$ width=35
\begin{scope}
\draw[fill=white] (23.500000, -0.000000) +(-45.000000:24.748737pt and 8.485281pt) -- +(45.000000:24.748737pt and 8.485281pt) -- +(135.000000:24.748737pt and 8.485281pt) -- +(225.000000:24.748737pt and 8.485281pt) -- cycle;
\clip (23.500000, -0.000000) +(-45.000000:24.748737pt and 8.485281pt) -- +(45.000000:24.748737pt and 8.485281pt) -- +(135.000000:24.748737pt and 8.485281pt) -- +(225.000000:24.748737pt and 8.485281pt) -- cycle;
\draw (23.500000, -0.000000) node {$\hat{K}(1/q)$};
\end{scope}
% Line 10: a b c e G $\text{DFT}_{q \times q}$ width=40
\draw (73.000000,60.000000) -- (73.000000,0.000000);
\begin{scope}
\draw[fill=white] (73.000000, 30.000000) +(-45.000000:28.284271pt and 50.911688pt) -- +(45.000000:28.284271pt and 50.911688pt) -- +(135.000000:28.284271pt and 50.911688pt) -- +(225.000000:28.284271pt and 50.911688pt) -- cycle;
\clip (73.000000, 30.000000) +(-45.000000:28.284271pt and 50.911688pt) -- +(45.000000:28.284271pt and 50.911688pt) -- +(135.000000:28.284271pt and 50.911688pt) -- +(225.000000:28.284271pt and 50.911688pt) -- cycle;
\draw (73.000000, 30.000000) node {$\text{DFT}_{q \times q}$};
\end{scope}
% Done with gates; drawing ending labels
\draw[color=black] (99.000000,60.000000) node[right] {${\langle 0 \vert}$};
\draw[color=black] (99.000000,45.000000) node[right] {${\vert \tilde{0}_L \rangle}$};
\draw[color=black] (99.000000,30.000000) node[right] {${\langle 0 \vert}$};
\draw[color=black] (99.000000,15.000000) node[anchor=mid west] {$\vdots$};
\draw[color=black] (99.000000,0.000000) node[right] {${\langle 0 \vert}$};
% Done with ending labels; drawing cut lines and comments
% Done with comments
\end{tikzpicture}
\vspace{2em}
\begin{tikzpicture}[scale=1.000000,x=1pt,y=1pt]
\filldraw[color=white] (0.000000, -7.500000) rectangle (99.000000, 67.500000);
% Drawing wires
% Line 1: a W {\vert \beta \rangle} {\vert \tilde{1}_L \rangle}
\draw[color=black] (0.000000,60.000000) -- (99.000000,60.000000);
\draw[color=black] (0.000000,60.000000) node[left] {${\vert \beta \rangle}$};
% Line 3: b W {\vert \beta \rangle} {\langle 0 \vert}
\draw[color=black] (0.000000,45.000000) -- (99.000000,45.000000);
\draw[color=black] (0.000000,45.000000) node[left] {${\vert \beta \rangle}$};
% Line 5: c W {\vert \beta \rangle} {\langle 0 \vert}
\draw[color=black] (0.000000,30.000000) -- (99.000000,30.000000);
\draw[color=black] (0.000000,30.000000) node[left] {${\vert \beta \rangle}$};
% Line 7: ...d W
\draw[color=black] (0.000000,15.000000) node[anchor=mid east] {$\vdots$};
% Line 8: e W {\vert \beta \rangle} {\langle 0 \vert}
\draw[color=black] (0.000000,0.000000) -- (99.000000,0.000000);
\draw[color=black] (0.000000,0.000000) node[left] {${\vert \beta \rangle}$};
% Done with wires; drawing gates
% Line 2: a G $\hat{K}(1/q)$ width=35
\begin{scope}
\draw[fill=white] (23.500000, 60.000000) +(-45.000000:24.748737pt and 8.485281pt) -- +(45.000000:24.748737pt and 8.485281pt) -- +(135.000000:24.748737pt and 8.485281pt) -- +(225.000000:24.748737pt and 8.485281pt) -- cycle;
\clip (23.500000, 60.000000) +(-45.000000:24.748737pt and 8.485281pt) -- +(45.000000:24.748737pt and 8.485281pt) -- +(135.000000:24.748737pt and 8.485281pt) -- +(225.000000:24.748737pt and 8.485281pt) -- cycle;
\draw (23.500000, 60.000000) node {$\hat{K}(1/q)$};
\end{scope}
% Line 4: b G $\hat{K}(1/q)$ width=35
\begin{scope}
\draw[fill=white] (23.500000, 45.000000) +(-45.000000:24.748737pt and 8.485281pt) -- +(45.000000:24.748737pt and 8.485281pt) -- +(135.000000:24.748737pt and 8.485281pt) -- +(225.000000:24.748737pt and 8.485281pt) -- cycle;
\clip (23.500000, 45.000000) +(-45.000000:24.748737pt and 8.485281pt) -- +(45.000000:24.748737pt and 8.485281pt) -- +(135.000000:24.748737pt and 8.485281pt) -- +(225.000000:24.748737pt and 8.485281pt) -- cycle;
\draw (23.500000, 45.000000) node {$\hat{K}(1/q)$};
\end{scope}
% Line 6: c G $\hat{K}(1/q)$ width=35
\begin{scope}
\draw[fill=white] (23.500000, 30.000000) +(-45.000000:24.748737pt and 8.485281pt) -- +(45.000000:24.748737pt and 8.485281pt) -- +(135.000000:24.748737pt and 8.485281pt) -- +(225.000000:24.748737pt and 8.485281pt) -- cycle;
\clip (23.500000, 30.000000) +(-45.000000:24.748737pt and 8.485281pt) -- +(45.000000:24.748737pt and 8.485281pt) -- +(135.000000:24.748737pt and 8.485281pt) -- +(225.000000:24.748737pt and 8.485281pt) -- cycle;
\draw (23.500000, 30.000000) node {$\hat{K}(1/q)$};
\end{scope}
% Line 9: e G $\hat{K}(1/q)$ width=35
\begin{scope}
\draw[fill=white] (23.500000, -0.000000) +(-45.000000:24.748737pt and 8.485281pt) -- +(45.000000:24.748737pt and 8.485281pt) -- +(135.000000:24.748737pt and 8.485281pt) -- +(225.000000:24.748737pt and 8.485281pt) -- cycle;
\clip (23.500000, -0.000000) +(-45.000000:24.748737pt and 8.485281pt) -- +(45.000000:24.748737pt and 8.485281pt) -- +(135.000000:24.748737pt and 8.485281pt) -- +(225.000000:24.748737pt and 8.485281pt) -- cycle;
\draw (23.500000, -0.000000) node {$\hat{K}(1/q)$};
\end{scope}
% Line 10: a b c e G $\text{DFT}_{q \times q}$ width=40
\draw (73.000000,60.000000) -- (73.000000,0.000000);
\begin{scope}
\draw[fill=white] (73.000000, 30.000000) +(-45.000000:28.284271pt and 50.911688pt) -- +(45.000000:28.284271pt and 50.911688pt) -- +(135.000000:28.284271pt and 50.911688pt) -- +(225.000000:28.284271pt and 50.911688pt) -- cycle;
\clip (73.000000, 30.000000) +(-45.000000:28.284271pt and 50.911688pt) -- +(45.000000:28.284271pt and 50.911688pt) -- +(135.000000:28.284271pt and 50.911688pt) -- +(225.000000:28.284271pt and 50.911688pt) -- cycle;
\draw (73.000000, 30.000000) node {$\text{DFT}_{q \times q}$};
\end{scope}
% Done with gates; drawing ending labels
\draw[color=black] (99.000000,60.000000) node[right] {${\vert \tilde{1}_L \rangle}$};
\draw[color=black] (99.000000,45.000000) node[right] {${\langle 0 \vert}$};
\draw[color=black] (99.000000,30.000000) node[right] {${\langle 0 \vert}$};
\draw[color=black] (99.000000,15.000000) node[anchor=mid west] {$\vdots$};
\draw[color=black] (99.000000,0.000000) node[right] {${\langle 0 \vert}$};
% Done with ending labels; drawing cut lines and comments
% Done with comments
\end{tikzpicture}
\caption{\label{fig:KDFTcat} The heralding of the logical cat states $\vert \tilde{0}_L \rangle$ (top) and $\vert \tilde{1}_L \rangle$ (bottom) for even $q$, where $q=2(L+1)$, performed in the limit of large amplitude.
The cat states are of amplitude $\alpha$, where $\beta = \alpha / \sqrt{q}$.}
\end{figure}

Fig.~\ref{fig:KDFTcat} shows two examples of this scheme in the even $q$ case, where by measuring vacuum on all modes but one, the heralded state will be---in the limit of large amplitude---equal to a $q$-component cat state $\vert \tilde{0}_L \rangle$ or $\vert \tilde{1}_L \rangle$.
For even $q$, where $q=2(L+1)$, we can choose to herald either logical codeword, as $\vert \tilde{1}_L \rangle$ is heralded on odd modes (counting the first mode as mode one), while the even modes herald $\vert \tilde{0}_L \rangle$.
Note that this exchange of codeword is limited to even $q$ because the codewords themselves are exclusively defined over even $q$ by Eqs.~(\ref{eqn:codewords0}) and (\ref{eqn:codewords1}), while for odd $q$ no matter the mode a state $\vert \text{cat}_q (\alpha) \rangle$ is heralded.
For our choice of logical basis, this ability to herald either codeword is advantageous given that the transformation between codewords (Pauli $X$) is otherwise a non-trivial number-translation operation, and only exact in the limit $|\alpha| \rightarrow \infty$ \cite{grimsmo2020quantum}.

At finite $|\alpha|$, the output is composed of a $q$-component cat state plus extraneous terms, and thus the fidelity is imperfect.
Prior to measurement, the terms of the cat state are associated exclusively with vacuum in the other modes, while for the extraneous terms these modes contain some amount of non-vacuum coherent light.
Thus for an all-vacuum measurement, the contributions from extraneous terms vanish as $|\alpha| \rightarrow \infty$.

\begin{figure*}[t]
\includegraphics{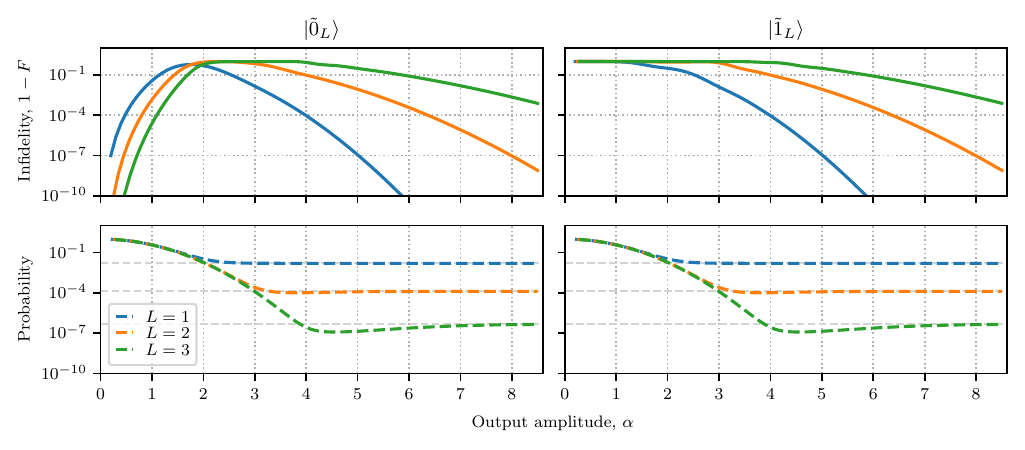}
\caption{\label{fig:qcat_probs} Infidelities and success probabilities for the states generated via our Kerr-DFT cat-generation protocol, with the top row showing infidelities to the logical codewords $\vert \tilde{0}_L \rangle$ and $\vert \tilde{1}_L \rangle$.
In the bottom row, where the success probabilities are shown, the grey, dashed horizontal lines indicate $q^{q-1}$, where $q=2(L+1)$.}
\end{figure*}

In Fig.~\ref{fig:qcat_probs} we show the probabilities and infidelities of generating the codewords $\vert \tilde{0}_L \rangle$ and $\vert \tilde{1}_L \rangle$.
For $\vert \tilde{0}_L \rangle$ the infidelity is very low at small amplitudes because the codeword itself approaches vacuum.
We also see that in all cases the success probability approaches $q^{q-1}$, which we show in Appendix~\ref{app:KDFT} to be related to the coefficients of the Kerr gate decomposition.

How probability scales with code order should be framed around the advantages of moving up in code order.
We can consider total probability over many sequential instances of state generation, such as in telecorrection, where increasing the order has been shown to reduce the required number of correction steps and in turn the number of ancillary states required \cite{boon2026generalised}.
Supposing we are in the limit of large amplitude, comparing $L=1$ to $L=3$, then after four stages $L=1$ has a lower probability of success than a single stage of $L=3$.

Alternatively, if one considers multiplexed state generation, then one would run some number of circuits at once to reach a target success probability.
If this is a target success probability of $90\%$, and again in the limit of large amplitude, then for $L=1$ and $L=3$ the number of multiplexed circuits to achieve this would be, respectively, $147$ versus $4\,828\,870$.

Another consideration to make is that, as-is, fidelity is directly tied to amplitude, which is apparent in Fig.~\ref{fig:qcat_probs}, meaning amplitude must increase to improve fidelity.
However, in practice, for a given amount of loss and specified performance target, there will be a single optimal amplitude at which the needed cat states should be prepared \cite{hastrup2022all, boon2026generalised}, with high amplitudes being problematic because they exacerbate the effects of loss.

In the following subsection, Sec.~\ref{sec:kdft_2}, we demonstrate the scheme analytically for the simplest case, $q=2$, showing both the vanishing of extraneous terms for large $|\alpha|$, and the ability to herald either $\vert \tilde{0}_0 \rangle$ or $\vert \tilde{1}_0 \rangle$ depending on the choice of mode.

\subsubsection{Kerr-DFT for $q=2$}
\label{sec:kdft_2}

For $q=2$ ($L=0$), our target states are
\begin{align}
\vert \tilde{0}_0 \rangle
=
\frac{\vert \alpha \rangle + \vert -\alpha \rangle}{\sqrt{\mathcal{N}_0}},
\quad
\vert \tilde{1}_0 \rangle
=
\frac{\vert \alpha \rangle - \vert -\alpha \rangle}{\sqrt{\mathcal{N}_1}}.
\end{align}
Prior to the application of the interferometer, our state is composed of two unentangled modes of Kerr states,
\begin{align}
\vert \psi \rangle
&=
\left[
\hat{K}(-1/2)
\vert \beta \rangle
\right]^{\otimes 2}
\\
&=
\left(
g_{0,1,2}
\vert \beta \rangle
+
g_{1,1,2}
\vert -\beta \rangle
\right)^{\otimes 2}
\\
&=
\frac{1}{2}
\bigl(
i
\vert \beta, \beta \rangle
+
\vert \beta, -\beta \rangle
+
\vert -\beta, \beta \rangle
-i
\vert -\beta, -\beta \rangle
\bigr).
\end{align}
With $W^{(2)}$ the DFT matrix of dimension $2$,
\begin{equation}
W^{(2)}
=
\frac{1}{\sqrt{2}}
\begin{pmatrix}
1 & 1 \\
1 & -1
\end{pmatrix},
\label{eqn:dft_2x2}
\end{equation}
we apply the DFT interferometer,
\begin{align}
\hat{W}^{(2)} \vert \psi \rangle
&=
\frac{1}{2}
\bigl(
i
\vert \alpha, 0 \rangle
+
\vert 0, \alpha \rangle
+
\vert 0, - \alpha \rangle
-i
\vert -\alpha, 0 \rangle
\bigr),
\end{align}
where $\alpha \equiv \sqrt{2} \beta$.

Measuring vacuum on the second mode heralds the state
\begin{align}
\langle 0 \vert_2
\hat{W}^{(2)} \vert \psi \rangle
&=
\frac{
\vert \alpha \rangle
-
\vert -\alpha \rangle
+
e^{-|\alpha|^2}
\vert 0 \rangle
}{2},
\end{align}
where we see that the final state approaches the codeword $\vert \tilde{1}_0 \rangle \propto \vert \alpha \rangle - \vert -\alpha \rangle$ for large $\alpha$,
\textit{i.e.},
\begin{equation}
\langle 0 \vert_2
\hat{W}^{(2)} \vert \psi \rangle
\propto
\vert \tilde{1}_0 \rangle
\quad
\text{for}
\quad
\alpha \rightarrow \infty,
\end{equation}
as the vacuum term vanishes, and the probability approaches $1/2$ given that $\mathcal{N}_1 \rightarrow 1/\sqrt{2}$.

Measuring vacuum on the first mode instead leads to perfect cancellation of the vacuum terms,
\begin{equation}
\langle 0 \vert_1 \hat{W}^{(2)} \vert \psi \rangle
=
\frac{
\vert \alpha \rangle
+
\vert -\alpha \rangle
}{2}
\propto
\vert \tilde{0}_0 \rangle,
\end{equation}
leading to an exact codeword at all amplitudes.
Although this perfect cancellation is not in general true for cases other than $q=2$, the choice of heralding mode can have an impact on fidelity at intermediate amplitudes, though the conditions in the limit of large amplitude always hold.
For $q$ in general we provide a derivation in Appendix~\ref{app:KDFT}.

\subsection{Bell cats from $q$-cats}
\label{sec:bell}

In order to perform telecorrection, one needs cat-encoded Bell states.
We now show how to transform two $q$-component cat states (such as from our Kerr-DFT scheme) into such entangled states.
Our scheme for the generation of cat-encoded Bell states is a generalisation to odd $L$ of the optical circuit proposed by Su et al. for $L=1$ \cite{su2022universal}.
To generate a cat-encoded Bell state of order $L$, where $q=2(L+1)$, one requires
\begin{itemize}
\item $2$ $q$-cat states of amplitude $\gamma$, where $\gamma = 2^{(L+1)/4} \alpha$,
\item $L+1$ modes of vacuum,
\item $3(L+1)/2$ beamsplitters given by Eq.~(\ref{eqn:dft_2x2}),
\item $L-1$ rotation gates,
\item $L+1$ photon-number-resolving detectors.
\end{itemize}
Fig.~\ref{fig:bell1} shows a variant of Su et al.'s circuit, whereas \ref{fig:bell5} shows its extension to $L=5$, illustrating the structure of our generalisation and its inclusion of rotation gates which are necessary for the generalisation.
Note that in general the structure of the Bell cat generation circuit involves two initially identical sets of modes, and five stages: initialisation, the mixing of modes within the sets, the application of rotations to one mode set, inter-set mixing of modes, and measurement.

The two sets of modes are initialised identically in the first stage: one mode with the resource state $\vert \text{cat}_q (\gamma)\rangle$, and $(L+1)/2$ modes of vacuum.
The second stage is also identical between these sets: the beamsplitters are used to mix the mode of the resource state with each mode of vacuum.
It is this mixing into the vacuum modes that leads to having to compensate in the initial amplitudes by the factor $2^{(L+1)/4}$ times that of the desired output amplitude.
Within these two stages, the transformations within both sets of modes are identical, and thus contain identical states.

In the third stage, rotations $\hat{R}(-j\pi/(2L+2))$, for $j=(1,\ldots,L)$, are applied to the last $L-1$ modes.
Note that this only impacts modes within the second mode set, and plays an essential role in allowing us to herald our desired state.
Subsequently, in the fourth stage, the two sets of vacuum-initialised modes are mixed: starting at mode $j=2$ and ending at mode $j=1+(L+1)/2$, ${50}{:}{50}$ beamsplitters mix each pair of modes $j$ and $j+1+(L+1)/2$.
Because of our rotations in stage three, the mixing of stage four produces modes of vacuum associated with the unwanted terms, which will allow us to exclude them via post-selection, \textit{i.e.}, a set of non-vacuum measurements will not contain these terms.

The final stage is to make photon-number measurements on all modes besides $j=0$ and $j=L-1$ (the modes which initially contained single-mode cat states), which succeeds in heralding a Bell cat if the measurements are all equal, $n$, and non-zero, $n \neq 0$.
The output state, if the input was perfect, will also be perfect (up to the deformation mentioned in Sec.~\ref{sec:cats} in our choice of logical basis).
We show an example encompassing all five stages in Appendix~\ref{app:bell3} for $L=3$, while Ref.~\cite{su2022universal} covers $L=1$.

Additionally, the parity of $n$ determines the sign of the Bell cat, $\vert \tilde{\Psi}^{\pm} \rangle$:
for $n$ odd, the output state will be positive, and for $n$ even, negative.
An optional rotation of $\hat{R}(\phi_L)$ may be applied to mode $j=L$ to convert the state to $\vert \tilde{\Phi}^{\pm} \rangle$, but this choice is of little importance to telecorrection, where teleporting with $\vert \tilde{\Phi} \rangle$ or $\vert \tilde{\Psi} \rangle$ only amounts to an exchange of Pauli operations applied to the output across the possible syndrome measurements.

\begin{figure}[t]
\includegraphics{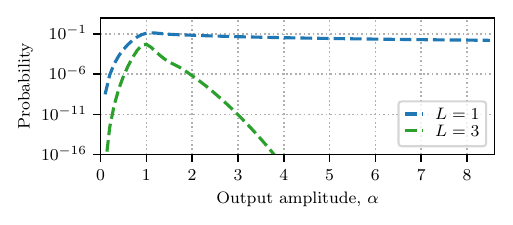}
\caption{\label{fig:bell_probs}
Success probabilities for the first two orders of our proposed the Bell cat generation protocol.
For $L=1$ the maximum success probability is ${\sim}{14.2\%}$, while for $L=3$ it is ${\sim}{0.530\%}$.}
\end{figure}

In Fig.~\ref{fig:bell_probs} we show the probability of success for our Bell cat generation protocol at $L=1$ and $L=3$.
Unlike our Kerr-DFT protocol for single-mode cat states, the success probability continues to decrease with amplitude, and is especially problematic as the order $L$ increases.
Due to how small the probability becomes at meaningful amplitudes for $L=3$, it is not numerically feasible to make comparisons like those made from the Kerr-DFT probabilities.

\begin{figure}[t]
\begin{tikzpicture}[scale=1.000000,x=1pt,y=1pt,fill=white]
\filldraw[color=white] (0.000000, -7.500000) rectangle (63.000000, 52.500000);
% Drawing wires
% Line 1: a W {\vert \text{cat}_4 (\gamma) \rangle}
\draw[color=black] (0.000000,45.000000) -- (63.000000,45.000000);
\draw[color=black] (0.000000,45.000000) node[left] {${\vert \text{cat}_4 (\gamma) \rangle}$};
% Line 2: b W {\vert 0 \rangle} {\langle n \vert}
\draw[color=black] (0.000000,30.000000) -- (49.500000,30.000000);
\draw[color=black] (0.000000,30.000000) node[left] {${\vert 0 \rangle}$};
% Line 5: c W {\vert \text{cat}_4 (\gamma) \rangle}
\draw[color=black] (0.000000,15.000000) -- (63.000000,15.000000);
\draw[color=black] (0.000000,15.000000) node[left] {${\vert \text{cat}_4 (\gamma) \rangle}$};
% Line 6: d W {\vert 0 \rangle} {\langle n \vert}
\draw[color=black] (0.000000,0.000000) -- (49.500000,0.000000);
\draw[color=black] (0.000000,0.000000) node[left] {${\vert 0 \rangle}$};
% Done with wires; drawing gates
% Line 3: a b
\draw (9.000000,45.000000) -- (9.000000,30.000000);
\filldraw (9.000000, 45.000000) circle(1.500000pt);
\filldraw (9.000000, 30.000000) circle(1.500000pt);
% Line 7: c d
\draw (9.000000,15.000000) -- (9.000000,0.000000);
\filldraw (9.000000, 15.000000) circle(1.500000pt);
\filldraw (9.000000, 0.000000) circle(1.500000pt);
% Line 9: b d
\draw (27.000000,30.000000) -- (27.000000,0.000000);
\filldraw (27.000000, 30.000000) circle(1.500000pt);
\filldraw (27.000000, 0.000000) circle(1.500000pt);
% Line 11: b END
\draw[color=black] (42.000000,30.000000) node[fill=white,right,minimum height=15.000000pt,minimum width=15.000000pt,inner sep=0pt] {\phantom{${\langle n \vert}$}};
\draw[color=black] (42.000000,30.000000) node[right] {${\langle n \vert}$};
% Line 12: d END
\draw[color=black] (42.000000,0.000000) node[fill=white,right,minimum height=15.000000pt,minimum width=15.000000pt,inner sep=0pt] {\phantom{${\langle n \vert}$}};
\draw[color=black] (42.000000,0.000000) node[right] {${\langle n \vert}$};
% Done with gates; drawing ending labels
% Done with ending labels; drawing cut lines and comments
% Done with comments
\end{tikzpicture}
\caption{\label{fig:bell1} An adapted version of the optical circuit from Ref.~\cite{su2022universal} for the generation of $L=1$ Bell cats.
Successful heralding is conditional on $n \neq 0$, and otherwise whether $n$ is odd or even determines whether the Bell cat is negative or positive.
The gates connecting wires, indicated by hollow barbells, are real, symmetric ${50}{:}{50}$  beamsplitters.}
\vspace{0.5cm}
\begin{tikzpicture}[scale=1.000000,x=1pt,y=1pt,fill=white]
\filldraw[color=white] (0.000000, -7.500000) rectangle (153.000000, 112.500000);
% Drawing wires
% Line 1: a W {\vert \text{cat}_{12} (\gamma) \rangle}
\draw[color=black] (0.000000,105.000000) -- (153.000000,105.000000);
\draw[color=black] (0.000000,105.000000) node[left] {${\vert \text{cat}_{12} (\gamma) \rangle}$};
% Line 2: b W {\vert 0 \rangle} {\langle n \vert}
\draw[color=black] (0.000000,90.000000) -- (139.500000,90.000000);
\draw[color=black] (0.000000,90.000000) node[left] {${\vert 0 \rangle}$};
% Line 3: c W {\vert 0 \rangle} {\langle n \vert}
\draw[color=black] (0.000000,75.000000) -- (139.500000,75.000000);
\draw[color=black] (0.000000,75.000000) node[left] {${\vert 0 \rangle}$};
% Line 4: d W {\vert 0 \rangle} {\langle n \vert}
\draw[color=black] (0.000000,60.000000) -- (139.500000,60.000000);
\draw[color=black] (0.000000,60.000000) node[left] {${\vert 0 \rangle}$};
% Line 9: f W {\vert \text{cat}_{12} (\gamma) \rangle}
\draw[color=black] (0.000000,45.000000) -- (153.000000,45.000000);
\draw[color=black] (0.000000,45.000000) node[left] {${\vert \text{cat}_{12} (\gamma) \rangle}$};
% Line 10: g W {\vert 0 \rangle} {\langle n \vert}
\draw[color=black] (0.000000,30.000000) -- (139.500000,30.000000);
\draw[color=black] (0.000000,30.000000) node[left] {${\vert 0 \rangle}$};
% Line 11: h W {\vert 0 \rangle} {\langle n \vert}
\draw[color=black] (0.000000,15.000000) -- (139.500000,15.000000);
\draw[color=black] (0.000000,15.000000) node[left] {${\vert 0 \rangle}$};
% Line 12: i W {\vert 0 \rangle} {\langle n \vert}
\draw[color=black] (0.000000,0.000000) -- (139.500000,0.000000);
\draw[color=black] (0.000000,0.000000) node[left] {${\vert 0 \rangle}$};
% Done with wires; drawing gates
% Line 5: a b
\draw (9.000000,105.000000) -- (9.000000,90.000000);
\filldraw (9.000000, 105.000000) circle(1.500000pt);
\filldraw (9.000000, 90.000000) circle(1.500000pt);
% Line 13: f g
\draw (9.000000,45.000000) -- (9.000000,30.000000);
\filldraw (9.000000, 45.000000) circle(1.500000pt);
\filldraw (9.000000, 30.000000) circle(1.500000pt);
% Line 6: a c
\draw (27.000000,105.000000) -- (27.000000,75.000000);
\filldraw (27.000000, 105.000000) circle(1.500000pt);
\filldraw (27.000000, 75.000000) circle(1.500000pt);
% Line 14: f h
\draw (27.000000,45.000000) -- (27.000000,15.000000);
\filldraw (27.000000, 45.000000) circle(1.500000pt);
\filldraw (27.000000, 15.000000) circle(1.500000pt);
% Line 7: a d
\draw (45.000000,105.000000) -- (45.000000,60.000000);
\filldraw (45.000000, 105.000000) circle(1.500000pt);
\filldraw (45.000000, 60.000000) circle(1.500000pt);
% Line 15: f i
\draw (45.000000,45.000000) -- (45.000000,0.000000);
\filldraw (45.000000, 45.000000) circle(1.500000pt);
\filldraw (45.000000, 0.000000) circle(1.500000pt);
% Line 17: TOUCH
% Line 19: b g
\draw (67.500000,90.000000) -- (67.500000,30.000000);
\filldraw (67.500000, 90.000000) circle(1.500000pt);
\filldraw (67.500000, 30.000000) circle(1.500000pt);
% Line 20: h G $\hat{R}_1$ width=15
\begin{scope}
\draw[fill=white] (67.500000, 15.000000) +(-45.000000:10.606602pt and 8.485281pt) -- +(45.000000:10.606602pt and 8.485281pt) -- +(135.000000:10.606602pt and 8.485281pt) -- +(225.000000:10.606602pt and 8.485281pt) -- cycle;
\clip (67.500000, 15.000000) +(-45.000000:10.606602pt and 8.485281pt) -- +(45.000000:10.606602pt and 8.485281pt) -- +(135.000000:10.606602pt and 8.485281pt) -- +(225.000000:10.606602pt and 8.485281pt) -- cycle;
\draw (67.500000, 15.000000) node {$\hat{R}_1$};
\end{scope}
% Line 21: TOUCH
% Line 22: c h
\draw (94.500000,75.000000) -- (94.500000,15.000000);
\filldraw (94.500000, 75.000000) circle(1.500000pt);
\filldraw (94.500000, 15.000000) circle(1.500000pt);
% Line 23: i G $\hat{R}_2$ width=15
\begin{scope}
\draw[fill=white] (94.500000, -0.000000) +(-45.000000:10.606602pt and 8.485281pt) -- +(45.000000:10.606602pt and 8.485281pt) -- +(135.000000:10.606602pt and 8.485281pt) -- +(225.000000:10.606602pt and 8.485281pt) -- cycle;
\clip (94.500000, -0.000000) +(-45.000000:10.606602pt and 8.485281pt) -- +(45.000000:10.606602pt and 8.485281pt) -- +(135.000000:10.606602pt and 8.485281pt) -- +(225.000000:10.606602pt and 8.485281pt) -- cycle;
\draw (94.500000, -0.000000) node {$\hat{R}_2$};
\end{scope}
% Line 24: d i
\draw (117.000000,60.000000) -- (117.000000,0.000000);
\filldraw (117.000000, 60.000000) circle(1.500000pt);
\filldraw (117.000000, 0.000000) circle(1.500000pt);
% Line 26: b c TOUCH
% Line 28: b END
\draw[color=black] (132.000000,90.000000) node[fill=white,right,minimum height=15.000000pt,minimum width=15.000000pt,inner sep=0pt] {\phantom{${\langle n \vert}$}};
\draw[color=black] (132.000000,90.000000) node[right] {${\langle n \vert}$};
% Line 29: c END
\draw[color=black] (132.000000,75.000000) node[fill=white,right,minimum height=15.000000pt,minimum width=15.000000pt,inner sep=0pt] {\phantom{${\langle n \vert}$}};
\draw[color=black] (132.000000,75.000000) node[right] {${\langle n \vert}$};
% Line 30: d END
\draw[color=black] (132.000000,60.000000) node[fill=white,right,minimum height=15.000000pt,minimum width=15.000000pt,inner sep=0pt] {\phantom{${\langle n \vert}$}};
\draw[color=black] (132.000000,60.000000) node[right] {${\langle n \vert}$};
% Line 32: g END
\draw[color=black] (132.000000,30.000000) node[fill=white,right,minimum height=15.000000pt,minimum width=15.000000pt,inner sep=0pt] {\phantom{${\langle n \vert}$}};
\draw[color=black] (132.000000,30.000000) node[right] {${\langle n \vert}$};
% Line 33: h END
\draw[color=black] (132.000000,15.000000) node[fill=white,right,minimum height=15.000000pt,minimum width=15.000000pt,inner sep=0pt] {\phantom{${\langle n \vert}$}};
\draw[color=black] (132.000000,15.000000) node[right] {${\langle n \vert}$};
% Line 34: i END
\draw[color=black] (132.000000,0.000000) node[fill=white,right,minimum height=15.000000pt,minimum width=15.000000pt,inner sep=0pt] {\phantom{${\langle n \vert}$}};
\draw[color=black] (132.000000,0.000000) node[right] {${\langle n \vert}$};
% Done with gates; drawing ending labels
% Done with ending labels; drawing cut lines and comments
% Done with comments
\end{tikzpicture}
\caption{\label{fig:bell5}
An optical circuit for $L=5$ Bell cat generation based on our generalisation of Fig.~\ref{fig:bell1}, with rotations $\hat{R}_j \equiv \hat{R}(-2j\pi/(L+1))$, and again conditional on $n \neq 0$.}
\end{figure}

\subsection{Single-mode damping}
\label{sec:damping}

For the purposes of telecorrection as described in Refs.~\cite{hastrup2022all, boon2026generalised}, it is necessary to prepare the Bell cat's first mode at a damped amplitude.
However the scheme we have described thus far does not work if the first and second resource cat states are of different amplitudes.
With an additional stage, one can instead damp the first mode of a prepared Bell state probabilistically---a procedure which works regardless of how one generated the initial state, and appears in the literature to model the finite envelope of GKP states \cite{tzitrin2020progress}.

\begin{figure}[t]
%! \usetikzlibrary{decorations.pathreplacing,decorations.pathmorphing}
\begin{tikzpicture}[scale=1.000000,x=1pt,y=1pt]
\filldraw[color=white] (0.000000, -7.500000) rectangle (27.000000, 37.500000);
% Drawing wires
% Line 1: a W {\vert 0 \rangle} {\langle 0 \vert}
\draw[color=black] (0.000000,30.000000) -- (27.000000,30.000000);
\draw[color=black] (0.000000,30.000000) node[left] {${\vert 0 \rangle}$};
% Line 2: b c W {\vert \tilde{\Psi}_L \rangle}<
\draw[color=black] (0.000000,15.000000) -- (27.000000,15.000000);
%   Deferring wire label at (0.000000,15.000000)
% Line 2: b c W {\vert \tilde{\Psi}_L \rangle}<
\draw[color=black] (0.000000,0.000000) -- (27.000000,0.000000);
\filldraw[color=white,fill=white] (0.000000,-3.750000) rectangle (-4.000000,18.750000);
\draw[decorate,decoration={brace,amplitude = 4.000000pt},very thick] (0.000000,-3.750000) -- (0.000000,18.750000);
\draw[color=black] (-4.000000,7.500000) node[left] {${\vert \tilde{\Psi}_L \rangle}$};
% Done with wires; drawing gates
% Line 3: a b G $\hat{U}_d$ width=15
\draw (13.500000,30.000000) -- (13.500000,15.000000);
\begin{scope}
\draw[fill=white] (13.500000, 22.500000) +(-45.000000:10.606602pt and 19.091883pt) -- +(45.000000:10.606602pt and 19.091883pt) -- +(135.000000:10.606602pt and 19.091883pt) -- +(225.000000:10.606602pt and 19.091883pt) -- cycle;
\clip (13.500000, 22.500000) +(-45.000000:10.606602pt and 19.091883pt) -- +(45.000000:10.606602pt and 19.091883pt) -- +(135.000000:10.606602pt and 19.091883pt) -- +(225.000000:10.606602pt and 19.091883pt) -- cycle;
\draw (13.500000, 22.500000) node {$\hat{U}_d$};
\end{scope}
% Done with gates; drawing ending labels
\draw[color=black] (27.000000,30.000000) node[right] {${\langle 0 \vert}$};
%   Deferring wire label at (27.000000,15.000000)
\filldraw[color=white,fill=white] (27.000000,-3.750000) rectangle (31.000000,18.750000);
\draw[decorate,decoration={brace,mirror,amplitude = 4.000000pt},very thick] (27.000000,-3.750000) -- (27.000000,18.750000);
\draw[color=black] (31.000000,7.500000) node[right] {${\vert \underline{\tilde{\Psi}}_L \rangle}$};
% Done with ending labels; drawing cut lines and comments
% Done with comments
\end{tikzpicture}
\caption{\label{fig:udamp} The damping protocol for Bell cat states, where the underline on the output indicates that the first mode has been damped.}
\end{figure}
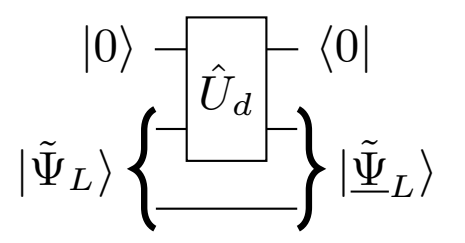

If a mode's amplitude need be damped from $\alpha$ to $\sqrt{1-\Gamma} \alpha$, then a beamsplitter of transmissivity $T=\sqrt{1-\Gamma}$ is required, given by
\begin{equation}
U_d
=
\begin{pmatrix}
\sqrt{1-\Gamma} & \sqrt{\Gamma} \\
\sqrt{\Gamma} & -\sqrt{1-\Gamma}
\end{pmatrix},
\end{equation}
which will mix one mode of the Bell state onto a mode of vacuum, as shown in Fig.~\ref{fig:udamp}.
After this operation, a measurement of vacuum on the new mode will produce the necessary damping, which can be accomplished with a threshold detector.
For an input $\vert \tilde{\Psi}_L \rangle$, this produces an output
\begin{equation}
\vert \underline{\tilde{\Psi}}_L \rangle
\propto
\frac{1}{\sqrt{2}}
\left(
\vert \underline{\tilde{0}}_L, \tilde{0}_L \rangle
+
\frac{\mathcal{N}_1}{\mathcal{N}_0}
\vert \underline{\tilde{1}}_L, \tilde{1}_L \rangle
\right),
\end{equation}
where the underbar indicates states of reduced amplitude,
\begin{equation}
\big\vert \tilde{J}_L \big\rangle,
=
\frac{1}{\sqrt{\mathcal{N}_1}}
\sum^{2 L + 1}_{m=0}
(-1)^{Jm}
\vert e^{i m \phi_L} \sqrt{1 - \Gamma} \alpha \rangle.
\end{equation}

As a vacuum measurement on a coherent state is ignorant to the state's phase, and the overall operation is independent of the other mode of the Bell state, we can demonstrate the effect with a single coherent state in one mode and vacuum in the second:
\begin{align}
\langle 0 \vert_1
\,
\hat{U}_d
\left(
\vert \alpha \rangle_0 \otimes \vert 0 \rangle_1
\right)
&=
\langle 0 \vert_1
\left(
\vert \sqrt{1-\Gamma} \alpha \rangle_0
\otimes
\vert \sqrt{\Gamma} \alpha \rangle_1
\right)
\notag
\\
&=
e^{-\Gamma|\alpha|^2/2}
\vert
\sqrt{1-\Gamma} \alpha
\rangle_0,
\label{eqn:damp_ex}
\end{align}
where modes are indicated in subscripts on the bras and kets.
Because the amount of loss per segment in telecorrection is generally very small, then as is the necessary damping, and thus in practice the impact on probability should be minor.
In fact we see in Eq.~(\ref{eqn:damp_ex}) that the probability should be ${\sim}{e^{-\Gamma|\alpha|^2}}$, which will be close to $1$ for $\Gamma$ sufficiently smaller than $|\alpha|$.

This could similarly be applied to single-mode cats to address the issues arising from fidelity being fixed to amplitude in the Kerr-DFT scheme, as in a high-amplitude cat state could be generated, then damped while maintaining its fidelity.
However the trends in fidelity for Kerr-DFT cats would demand large amounts of damping to see substantial improvements, in which case the success probabilities will be inherently low.
We would also not expect the damping operation to be itself perfect, and thus may be introduce more error than compensated for by the theoretical increase in fidelity.

\section{Deterministic cat generation}
\label{sec:deterministic}

The techniques we have proposed so far have relied on access to the Kerr gate, but have otherwise employed interferometry and post-selection, where the fidelity to a cat state is determined by the state's amplitude.
While we benefit from the number of gates involved being fixed per code order and agnostic to our target fidelities, post-selection in particular means that state generation can quickly become low-rate at higher orders, or will otherwise massively inflate the resource costs to produce states at a desired rate.
Bell cat states in particular are problematic, as they have vanishingly small probabilities as amplitude increases.
However, as we already assume access to Kerr gates, then other techniques are available to us without relying on post-selection, allowing us to theoretically perform state generation deterministically.

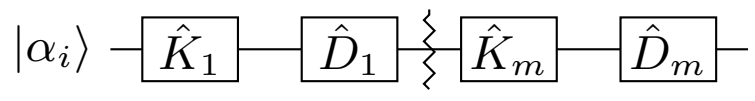
\begin{figure}[t]
%! \usetikzlibrary{decorations.pathreplacing,decorations.pathmorphing}
\begin{tikzpicture}[scale=1.000000,x=1pt,y=1pt]
\filldraw[color=white] (0.000000, -7.500000) rectangle (120.000000, 7.500000);
% Drawing wires
% Line 1: a W {\vert \alpha_i \rangle}
\draw[color=black] (0.000000,0.000000) -- (120.000000,0.000000);
\draw[color=black] (0.000000,0.000000) node[left] {${\vert \alpha_i \rangle}$};
% Done with wires; drawing gates
% Line 2: a G $\hat{K}_1$ width=18
\begin{scope}
\draw[fill=white] (15.000000, -0.000000) +(-45.000000:12.727922pt and 8.485281pt) -- +(45.000000:12.727922pt and 8.485281pt) -- +(135.000000:12.727922pt and 8.485281pt) -- +(225.000000:12.727922pt and 8.485281pt) -- cycle;
\clip (15.000000, -0.000000) +(-45.000000:12.727922pt and 8.485281pt) -- +(45.000000:12.727922pt and 8.485281pt) -- +(135.000000:12.727922pt and 8.485281pt) -- +(225.000000:12.727922pt and 8.485281pt) -- cycle;
\draw (15.000000, -0.000000) node {$\hat{K}_1$};
\end{scope}
% Line 3: a G $\hat{D}_1$ width=18
\begin{scope}
\draw[fill=white] (45.000000, -0.000000) +(-45.000000:12.727922pt and 8.485281pt) -- +(45.000000:12.727922pt and 8.485281pt) -- +(135.000000:12.727922pt and 8.485281pt) -- +(225.000000:12.727922pt and 8.485281pt) -- cycle;
\clip (45.000000, -0.000000) +(-45.000000:12.727922pt and 8.485281pt) -- +(45.000000:12.727922pt and 8.485281pt) -- +(135.000000:12.727922pt and 8.485281pt) -- +(225.000000:12.727922pt and 8.485281pt) -- cycle;
\draw (45.000000, -0.000000) node {$\hat{D}_1$};
\end{scope}
% Line 4: a BARRIER
\draw[decorate,decoration={zigzag,amplitude=1pt,segment length=4}] (60.000000,-7.500000) -- (60.000000,7.500000);
% Line 5: a G $\hat{K}_m$ width=18
\begin{scope}
\draw[fill=white] (75.000000, -0.000000) +(-45.000000:12.727922pt and 8.485281pt) -- +(45.000000:12.727922pt and 8.485281pt) -- +(135.000000:12.727922pt and 8.485281pt) -- +(225.000000:12.727922pt and 8.485281pt) -- cycle;
\clip (75.000000, -0.000000) +(-45.000000:12.727922pt and 8.485281pt) -- +(45.000000:12.727922pt and 8.485281pt) -- +(135.000000:12.727922pt and 8.485281pt) -- +(225.000000:12.727922pt and 8.485281pt) -- cycle;
\draw (75.000000, -0.000000) node {$\hat{K}_m$};
\end{scope}
% Line 6: a G $\hat{D}_m$ width=18
\begin{scope}
\draw[fill=white] (105.000000, -0.000000) +(-45.000000:12.727922pt and 8.485281pt) -- +(45.000000:12.727922pt and 8.485281pt) -- +(135.000000:12.727922pt and 8.485281pt) -- +(225.000000:12.727922pt and 8.485281pt) -- cycle;
\clip (105.000000, -0.000000) +(-45.000000:12.727922pt and 8.485281pt) -- +(45.000000:12.727922pt and 8.485281pt) -- +(135.000000:12.727922pt and 8.485281pt) -- +(225.000000:12.727922pt and 8.485281pt) -- cycle;
\draw (105.000000, -0.000000) node {$\hat{D}_m$};
\end{scope}
% Done with gates; drawing ending labels
% Done with ending labels; drawing cut lines and comments
% Done with comments
\end{tikzpicture}
\caption{\label{fig:dk_stages}
The first and final stages of the $m$-stage circuit we used for the deterministic conversion of a coherent state into approximate cat states, where $\hat{K}_j \equiv \hat{K}(\kappa_j)$ and $\hat{D}_j \equiv \hat{D}(\beta_j)$.
The initial coherent state can also be thought of as a displacement of $\alpha_i = \beta_0$ acting on vacuum.}
\end{figure}

On account of the universality of displacement combined with the Kerr gate, schemes similar to Fig.~\ref{fig:dk_stages} have been proposed for optical state generation (sometimes with the inclusion of other Gaussian operations), having been used for Fock, GKP, and other highly-structured Bosonic states \cite{meher2026generation, blair2025towards, fur2026deterministic, rudolph2026physical}.
With sufficiently many stages and an appropriate choice of parameters, it should be possible to generate any state this way up to some target fidelity, irrespective of amplitude.
Moreover, the structural similarity between the decomposed Kerr gates and each order of the cat code suggests that this type of scheme may be particularly well-suited to generating cat states.

We find that by using the circuit of Fig.~\ref{fig:dk_stages} while keeping the Kerr gates fixed to the rational parameters ${\pm}{1/q}$, where $q$ is the number of components in our target cat state, we can derive displacements which generate high-fidelity cat states in relatively few stages, and do this with much lower numerical overhead compared to optimisation where $\hat{K}$ is kept entirely free.
This simplified approach has also allowed us to better understand the mechanics behind how this scheme works with respect to cat states, and revealed characteristics that make the techniques for cat generation more robust and generalisable.

\subsection{Inline Kerr-displacement $q$-cats}
\label{sec:inline_qcats}

\subsubsection{Single-stage generation}
\label{sec:dk_single}

\begin{figure}[t]
\includegraphics{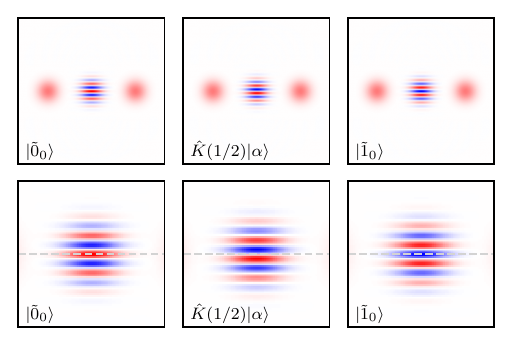}
\caption{\label{fig:2wigs} The Wigner functions of the indicated $q=2$ states, with the bottom row showing a close-up of the fringes.
All states are prepared at amplitude $\alpha=3.0$.}
\end{figure}

The simplest case to imagine for the scheme in Fig.~\ref{fig:dk_stages} is one where $m=1$.
In this case, with the Kerr gate restricted to $\hat{K}(1/q)$
and the initial amplitude taken to be equal to the target amplitude, $\alpha_i = \alpha_t$, then our only free variable is a single displacement, and we may write the state as
\begin{equation}
\hat{D}(\beta) \hat{K}(1/q) \vert \alpha_t \rangle.
\end{equation}
With this in mind, consider the fringes of the Wigner functions shown in Fig.~\ref{fig:2wigs} for the two-component cat states (left and right) and the two-component Kerr state (centre).
The amplitudes of the fringes of the Kerr state compared to those of the cat states are different only in phase, and thus displacements along that axis will bring us approximately close to either cat state.

\begin{figure}[t]
\includegraphics{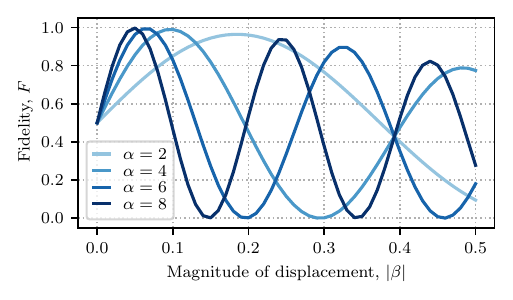}
\caption{\label{fig:single_stage_twocat} Fidelities for two-component cat states prepared using a single stage of displacement and Kerr gates, given by $\hat{D}(\beta)\hat{K}(1/2)\vert \alpha_t \rangle$.
This was evaluated for various amplitudes, $\alpha$, scanned over magnitudes of displacement for $i\beta \in \mathbb{R}$.}
\end{figure}

In Fig.~\ref{fig:single_stage_twocat}, we see the fidelities achieved this way for a span of displacements, and at various amplitudes for the two-component cat states.
As the amplitude is increased, higher maximum fidelities become possible, and occur at lower displacements, as the distance between fringes becomes small relative to the size of the two lobes of the cat state in phase space---displacements mis-align the lobes, and thus the smaller the displacement to align the fringes the less the lobe mis-alignment.
More specifically, the distance between fringes is inversely proportional to the state's amplitude, and thus so is our ideal displacement.
We also observe damped periodic behaviour
at multiples of the ideal displacement,
as one would expect from matching the fringes but further mis-aligning the lobes.
The product of the ideal displacement times target amplitude becomes approximately fixed as amplitude increases,
allowing one to determine ideal displacements via a scaling parameter.

\begin{figure}[t]
\includegraphics{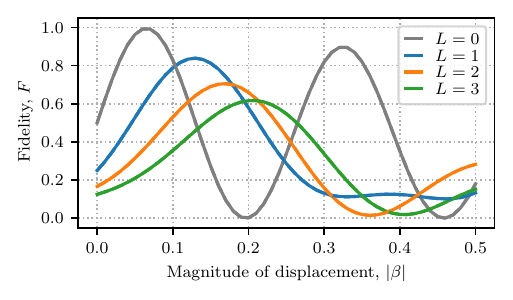}
\caption{\label{fig:single_stage} The fidelity of single-stage Kerr-displacement cat generation, $\hat{D}(\beta)\hat{K}(1/q)\vert \alpha_t \rangle$ for $q = 2(L+1)$ and $i\beta \in \mathbb{R}$.
Solid lines indicate fidelity to $\vert \tilde{0}_L \rangle$, and dashed indicate $\vert \tilde{1}_L \rangle$ (which are nearly identical to their logical counterpart in all cases).
Target states were of amplitude $\alpha_t=6.0$, and the maximum fidelities for $L=\{0,1,2,3\}$ are $F\approx\{0.992, 0.839, 0.706, 0.617\}$ for both codewords.}
\end{figure}

This single-stage technique can similarly be applied to higher-order cat states, although the increasing degree of rotational symmetry reduces how accurate the approximation by a single displacement can be.
In Fig.~\ref{fig:single_stage} we show the fidelities achieved this way, where the fidelities to $\vert \tilde{0}_L \rangle$ and $\vert \tilde{1}_L \rangle$ are approximately identical up to a sign-flip in the displacement.
Although the approximation falls off quickly as the number of components rises, the notions of displacement scaling and codeword-switching will prove useful later for the many-stage cases.

In the following section we use the decomposition of the Kerr gate to demonstrate the effects behind single-stage two-component cat generation analytically.
We will show both the effect of displacement scaling, and how a sign-flip on the displacement can take us between approximations of $\vert \tilde{0}_0 \rangle$ and $\vert \tilde{1}_0 \rangle$.

\subsubsection{Single-stage $q=2$ derivation}

Here via a single stage of Kerr-displacement we will demonstrate the generation of the $q=2$ ($L=0$) cat state,
\begin{equation}
\vert \text{cat}_2 (\alpha) \rangle
=
\vert \tilde{0}_0 \rangle
=
\frac{1}{\sqrt{\mathcal{N}_0}}
\left(
\vert \alpha \rangle + \vert -\alpha \rangle
\right),
\end{equation}
leaving the order on $\vert \tilde{0} \rangle$ implicit from here on.
Generation of this state begins with a coherent state of amplitude $\alpha \in \mathbb{R}$, applying a Kerr gate $\hat{K}(1/2)$, where
\begin{equation}
\hat{K}(1/2)
=
e^{i \pi/4}
\hat{R}(0)
+
e^{-i \pi/4}
\hat{R}(-\pi),
\end{equation}
then a displacement $\hat{D}(\beta)$, where $i \beta \in \mathbb{R}$,
\begin{align}
&\hat{D}(\beta) \hat{K}(1/2) \vert \alpha \rangle
\notag
\\
&=
\hat{D}(\beta)
\frac{1}{\sqrt{2}}
\left(
e^{i \pi /4} \vert \alpha \rangle
+
e^{-i \pi /4} \vert -\alpha \rangle
\right)
\\
&=
\frac{1}{\sqrt{2}}
\left(
e^{i \pi /4 + \beta \alpha} \vert \alpha + \beta \rangle
+
e^{-i \pi /4 - \beta \alpha} \vert -\alpha + \beta \rangle
\right),
\label{eqn:dk}
\end{align}
where for our choice of $\alpha$ and $\beta$, using Eq.~(\ref{eqn:displacement}), we have
\begin{align}
\hat{D}(\beta) \vert \alpha \rangle
&=
e^{(\beta \alpha^* - \beta^* \alpha)/2}
\vert \alpha + \beta \rangle
=
e^{\beta \alpha}
\vert \alpha + \beta \rangle.
\end{align}
From here consider that our aim is to maximise the overlap of this state with $\vert \tilde{0} \rangle$, which is given by
\begin{align}
&\langle \tilde{0} \vert \hat{D}(\beta) \hat{K}(1/2) \vert \alpha \rangle
\notag
\\
&=
\frac{1}{\sqrt{2 \mathcal{N}_0}}
\sum^{ \{ 0,1 \} }_{j,k}
e^{(-1)^j \left( i \pi/4 + \beta \alpha \right)}
\langle (-1)^j \alpha \vert (-1)^k \alpha + \beta \rangle,
\label{eqn:cat_overlap}
\end{align}
from which, using Eq.~(\ref{eqn:coherent_overlap}), we may write
\begin{align}
\langle \tilde{0} \vert \hat{D}(\beta) \hat{K}(1/2) \vert \alpha \rangle
\notag
\\
=
\frac{1}{\sqrt{2 \mathcal{N}_0}}
\biggl(
& e^{i \pi/4}
\bigl[
e^{2 \beta \alpha - | \beta |^2/2}
+
e^{-| 2\alpha + \beta |^2/2}
\bigr]
+
\notag
\\
& e^{-i \pi/4}
\bigl[
e^{-| 2\alpha + \beta |^2/2}
+
e^{-2 \beta \alpha - | \beta |^2/2}
\bigr]
\biggr).
\label{eqn:overlap_full}
\end{align}

At this point we will make the approximation $\alpha \gg \pi/4 \gg |\beta|$.
As we saw previously, the displacement must be small relative to the amplitude of the initial coherent state, otherwise there will not be strong overlap between the lobes.
We also wish to operate in a regime where $\alpha$ is large generally, as for $\alpha \rightarrow 0$ our target cat state is equal to vacuum, which, while the overlap may be maximised, is not useful.
Using our approximation, we find
\begin{align}
&\langle \tilde{0} \vert \hat{D}(\beta) \hat{K}(1/2) \vert \alpha \rangle
\notag
\\
&\approx
\frac{1}{\sqrt{2 \mathcal{N}_0}}
\left(
{\exp}{\left[ i \frac{\pi}{4} + 2 \beta \alpha \right]}
+
{\exp}{\left[ - i \frac{\pi}{4} - 2 \beta \alpha \right]}
\right).
\label{eqn:overlap_approx}
\end{align}
If we write this in terms of the fidelity, we have,
\begin{align}
\vert
\langle \tilde{0} \vert \hat{D}( \beta ) &\hat{K}(1/2) \vert \alpha \rangle
\vert^2
\notag
\\
&\approx
\frac{1}{2 \mathcal{N}_0}
\left(
2
+
e^{i \pi/2} e^{4 \beta \alpha}
+
e^{-i \pi/2} e^{-4 \beta \alpha}
\right)
\\
&=
\frac{1}{2 \mathcal{N}_0}
\left(
2
+
\frac{i}{\sqrt{2}}
\left[
e^{4 \beta \alpha}
-
e^{-4 \beta \alpha}
\right]
\right)
\\
&=
\frac{1}{2 \mathcal{N}_0}
\left(
2
\left[
1
-
\frac{{\sin}{(4 b \alpha)}}{\sqrt{2}}
\right]
\right),
\label{eqn:q2_F0}
\end{align}
having taken $b = i \beta$, where we see that the condition to satisfy is
\begin{equation}
b \alpha = - \frac{\pi}{8} \approx - 0.393,
\end{equation}
and implies that the ideal $\beta$ decreases with increasing $\alpha$, which aligns with our numerical results. 
Numerically, for $\alpha = 8$, we found that fidelity was maximised at $b_{\text{num}} = -0.05$ (up to the precision of steps $0.01$ in $b_{\text{num}}$), \textit{i.e.},
\begin{equation}
b_{\text{num}} \alpha = - 0.4.
\end{equation}
The condition is also evidently periodic, but we expect the approximation to break down for increasing $\beta$ as we derived our expression under the assumption that $\alpha \gg \beta$.

We can also observe the effect of flipping the sign of the displacement, and how this takes us to an approximately equal fidelity to $\vert \tilde{1}_0 \rangle$.
In doing so we find
\begin{align}
\vert
\langle \tilde{1} \vert \hat{D}( -\beta ) &\hat{K}(1/2) \vert \alpha \rangle
\vert^2
\notag
\\
&\approx
\frac{1}{2 \mathcal{N}_1}
\left(
2
-
e^{i \pi/2} e^{-4 \beta \alpha}
-
e^{-i \pi/2} e^{4 \beta \alpha}
\right)
\\
&=
\frac{1}{2 \mathcal{N}_1}
\left(
2
+
\frac{i}{\sqrt{2}}
\left[
e^{4 \beta \alpha}
-
e^{-4 \beta \alpha}
\right]
\right)
\\
&=
\frac{1}{2 \mathcal{N}_1}
\left(
2
\left[
1
-
\frac{{\sin}{(4 b \alpha)}}{\sqrt{2}}
\right]
\right),
\label{eqn:q2_F1}
\end{align}
where Eq.~(\ref{eqn:q2_F1}) is equal to Eq.~(\ref{eqn:q2_F0}) up to the normalisation constants.
However $\mathcal{N}_0 \rightarrow \mathcal{N}_1$ with increasing $\alpha$, so the two cases are approximately equal for sufficiently large amplitudes.

\subsubsection{Multi-stage generation}
\label{sec:dk_multi}

\begin{figure}[t]
\includegraphics{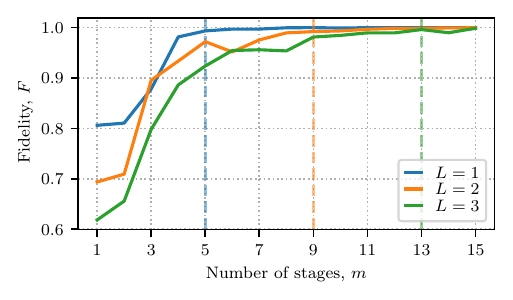}
\caption{\label{fig:opt_fidelities} Fidelities to $\vert \tilde{0}_L \rangle$ achieved by numerically optimising the initial amplitude and displacements for a given number of stages and at each indicated order of the cat code.
For $L~=~\{1,2,3\}$, the target amplitudes of the cat states were $\alpha_t~=~\{3.2, 4.7, 6.0\}$.
Vertical dashed lines indicate the first cases exceeding $99.0\%$ fidelity.}
\end{figure}

For multi-stage state generation, we again kept the Kerr gates fixed to $\hat{K}(1/q)$, but used the software \texttt{MrMustard}~\cite{mrmustard} to simulate the optical circuits in Fock space then numerically optimise the initial amplitudes and each displacement via gradient descent.
To generate an approximate $q$-component cat state of amplitude $\alpha_t$, we used the circuit of Fig.~\ref{fig:dk_stages} with the following components:
\begin{itemize}
\item a single coherent state of amplitude $\alpha_i$,
\item $m$ Kerr gates, $\hat{K}(1/q)$,
\item $m$ displacement gates, $\hat{D}(\beta_j)$.
\end{itemize}
When optimising, we initialised our free parameters at $\alpha_i = \alpha_t$ and $\beta_j = 0$.
Although one must nominally re-optimise for any change in the target cat state, we will describe two useful effects generalised from our results at $m=1$: one related to the target amplitude and the other to the target logical codeword, both controlled by the displacements, and both of which can be taken advantage of to reduce numerical overhead.
Additionally, although we will focus on even $q$ here for brevity and to highlight effects related to our choice of cat basis, the fundamentals generalise to odd $q$ as well.

In Fig.~\ref{fig:opt_fidelities} we show the fidelities found via numerical optimisation for a given number of stages, stopping when a threshold of $99.0\%$ fidelity was exceeded;
note that the target amplitudes were not identical,
but chosen such that inter-term overlaps were sufficiently small, as working in this regime will be useful later.
Although our results here were achieved only with numerical optimisation and do not necessarily represent globally optimal parameters, there are clear trends:
fidelity improved as the number of stages increased, but the approach towards unit fidelity appears asymptotic, and the number of stages required to reach a given fidelity generally grew with code order.
Considering this in terms of Kerr gates, this differs from our Kerr-DFT scheme, where the fidelity is determined solely by the involved amplitudes while the number of Kerr gates is fixed by the code order.

\begin{figure*}[t]
\includegraphics{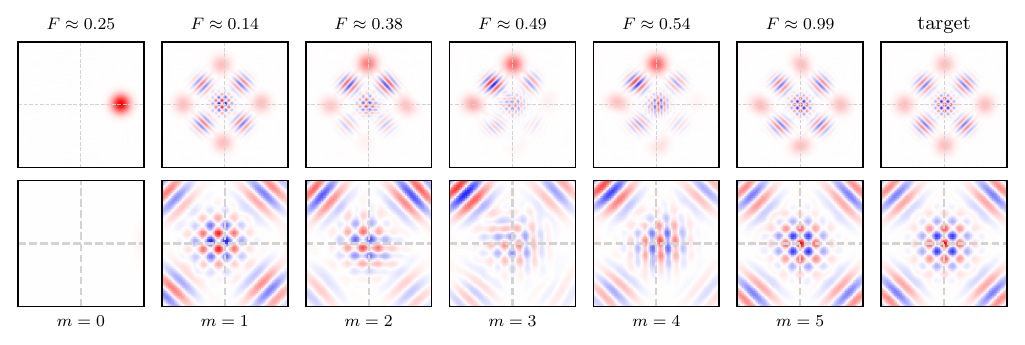}
\caption{\label{fig:wig_stages} Stages of the generation of an approximate $\vert \tilde{0}_1 \rangle$, where $\alpha = 3.2$, where each plot (with the bottom row zoomed in on the origin) shows the Wigner function of the state after the application of a single stage, consisting of a Kerr gate $\hat{K} \equiv \hat{K}(1/4)$ and the displacement $\hat{D}(\beta_j)$, using the parameters which produce the results for $L=1$ and $m=5$ in Fig.~\ref{fig:opt_fidelities}.
The displacements were, inclusive of the initial coherent state, $\beta_j = \{ 3.096-0.056i,\, -0.194-0.083i,\, 0.071-0.068i,\, 0.014-0.107i,\, -0.029-0.023i,\, -0.145+0.071i \}$.}
\end{figure*}

As an example of the evolution of the state, in Fig.~\ref{fig:wig_stages} we show the Wigner function at each stage of combined Kerr and displacement gates for $L=1$ and $m=5$;
at $m=0$ the coherent state at $\alpha_i$ has been prepared, while $m=1$ shows $\hat{D}(\beta_1) \hat{K}(1/4) \vert \alpha_i \rangle$, and so on.
Although our intuition from the $q=2$, $m=1$ case is that fringe alignment is essential, we can see that as we introduce more stages the effect of each stage bares a less obvious relationship to the fringes.
While the Kerr gates are essential to the functioning of this method of state generation, at a given stage they do not have a significant impact on fidelity: the fidelity remains approximately unchanged when the Kerr gate of any stage is applied.

\begin{figure}[t]
\includegraphics{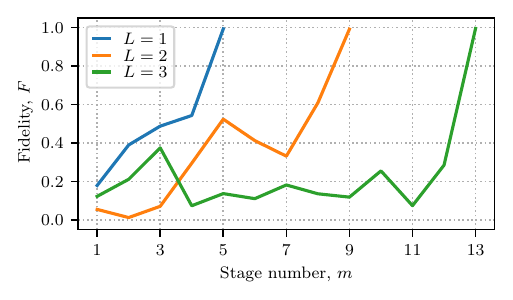}
\caption{\label{fig:fidelity_stages} Fidelities at each stage of the single-mode Kerr-displacement protocol, where each stage includes a Kerr gate and displacement, with the parameters corresponding to the first cases which produce the ${>}{99.0}\%$ fidelities presented in Fig.~\ref{fig:opt_fidelities}.}
\end{figure}

It is also worth noting that based on the displacements we found numerically, improvement at each stage is non-monotonic: to achieve the fidelities we demonstrated, the state must often go through a series of stages that first reduce its fidelity to the target cat state.
This can be observed in Fig.~\ref{fig:fidelity_stages}.
This progression increases the cost of numerical optimisation, as it cannot be performed sequentially across the stages, and thus the search space grows exponentially with the number of stages rather than linearly.
However we will now elaborate on the two techniques we found, which may be used to reduce numerical overhead when one wishes to generate cat states of varying parameters.

\begin{figure}[t]
\includegraphics{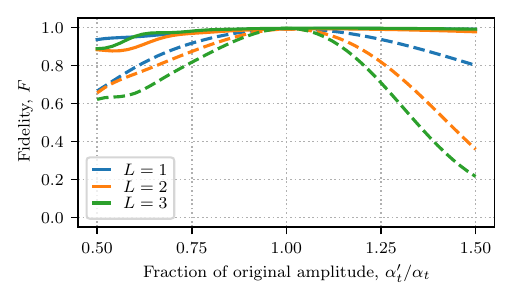}
\caption{\label{fig:param_scaling} Fidelities found at different amplitudes using parameters derived from those producing the maxima of Fig.~\ref{fig:opt_fidelities}.
Solid: the initial amplitudes and displacements are scaled according to Eqs.~(\ref{eqn:i_scale}) and (\ref{eqn:d_scale}), respectively.
Dashed: only the initial amplitude is scaled according to Eq.~(\ref{eqn:i_scale}).
Not shown: if the displacements are scaled as $\beta'_j = \beta_j (\alpha'_t / \alpha_t)$ the performance is significantly worse than both of the prior cases, and so we have opted to exclude those results for clarity.}
\end{figure}

The first effect that can be taken advantage of is a set of scaling rules for the parameters, similar to what was noted in the prior section for $m=1$.
From values optimised for a target amplitude of $\alpha_t$, to scale them to a new target amplitude $\alpha'_t$, the process is as follows:
if the initial coherent state was prepared at $\alpha_i$, the new value $\alpha'_i$ is given by
\begin{equation}
\alpha'_i
=
\alpha_i \frac{\alpha'_t}{\alpha_t},
\label{eqn:i_scale}
\end{equation}
while for displacements $\beta_j$, the scaled values $\beta'_j$ are given by
\begin{equation}
\beta'_j
=
\beta_j \frac{\alpha_t}{\alpha'_t},
\label{eqn:d_scale}
\end{equation}
where the relationship to the cat state amplitudes is the inverse.
Fig.~\ref{fig:param_scaling} shows the fidelities achieved this way compared to only scaling the initial coherent state amplitudes $\alpha_i$, where there is a dramatic improvement particularly at higher amplitudes from following both scaling rules, while the effect is weaker at lower amplitudes.

\begin{figure}[t]
\includegraphics{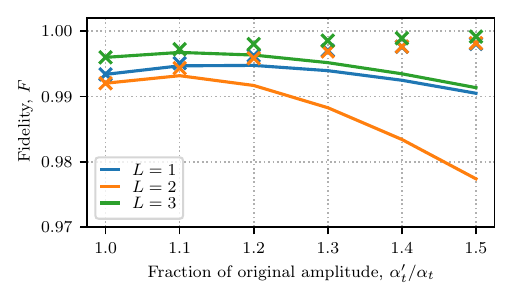}
\caption{\label{fig:scale_opt} Fidelities of states generated via rescaled displacements (solid) and rescaled displacements that were further optimised numerically (crosses).}
\end{figure}

These scaled parameters are not, however, the best that can be achieved even just numerically, but for higher amplitudes can be used to produce seed parameters from which to optimise, as shown in Fig.~\ref{fig:scale_opt}.
This potentially reduces the resources required for numerical optimisation, which is especially important at larger numbers of stages and for increasing amplitudes, as increasing amplitude will drive up the necessary Fock cutoff, in turn making the simulation of each operation more expensive.

The second useful effect we encountered generalises the sign-flip effect from $m=1$:
although Fig.~\ref{fig:opt_fidelities} only considers target states of $\vert \tilde{0}_L \rangle$, we found that flipping the sign on every odd or even displacement (where the initial coherent state amplitude is also included) produces a state $\vert \tilde{1}_L \rangle$ of approximately equal fidelity.
This holds for any $\alpha_i \approx \alpha_t$ and $\beta_j \ll \alpha_i$, \textit{i.e.}, not only for optimised displacements.
In our scheme, the odd case can be written as
\begin{align}
&\left\vert
\langle
\tilde{0}
\vert
\left(
\prod^{m}_{j=1}
\hat{D}(\beta_j)
\hat{K}(1/q)
\right)
\vert
\alpha_i
\rangle
\right\vert^2
\notag
\\
&\approx
\left\vert
\langle
\tilde{1}
\vert
\left(
\prod^{m}_{j=1}
\hat{D}((-1)^j \beta_j)
\hat{K}(1/q)
\right)
\vert
\alpha_i
\rangle
\right\vert^2,
\end{align}
and the even case similarly (with a flip on $\alpha_i$); here $\vert \tilde{0} \rangle$ and $\vert \tilde{1} \rangle$ are even-$q$ cat states of any order.
In Appendix~\ref{app:pk} we demonstrate the mechanism behind this analytically by investigating how the sign-flips are equivalent to the parity operator acting on the Kerr gates.

\subsection{Inline Bell cats}
\label{sec:dku}

\begin{figure}[t]
%! \usetikzlibrary{decorations.pathreplacing,decorations.pathmorphing}
\begin{tikzpicture}[scale=1.000000,x=1pt,y=1pt]
\filldraw[color=white] (0.000000, -7.500000) rectangle (192.000000, 22.500000);
% Drawing wires
% Line 1: a W {\vert \alpha_{0,i} \rangle}
\draw[color=black] (0.000000,15.000000) -- (192.000000,15.000000);
\draw[color=black] (0.000000,15.000000) node[left] {${\vert \alpha_{0i} \rangle}$};
% Line 2: b W {\vert \alpha_{1,i} \rangle}
\draw[color=black] (0.000000,0.000000) -- (192.000000,0.000000);
\draw[color=black] (0.000000,0.000000) node[left] {${\vert \alpha_{1i} \rangle}$};
% Done with wires; drawing gates
% Line 3: a G $\hat{K}_{01}$ width=21
\begin{scope}
\draw[fill=white] (16.500000, 15.000000) +(-45.000000:14.849242pt and 8.485281pt) -- +(45.000000:14.849242pt and 8.485281pt) -- +(135.000000:14.849242pt and 8.485281pt) -- +(225.000000:14.849242pt and 8.485281pt) -- cycle;
\clip (16.500000, 15.000000) +(-45.000000:14.849242pt and 8.485281pt) -- +(45.000000:14.849242pt and 8.485281pt) -- +(135.000000:14.849242pt and 8.485281pt) -- +(225.000000:14.849242pt and 8.485281pt) -- cycle;
\draw (16.500000, 15.000000) node {$\hat{K}_{01}$};
\end{scope}
% Line 4: b G $\hat{K}_{11}$ width=21
\begin{scope}
\draw[fill=white] (16.500000, -0.000000) +(-45.000000:14.849242pt and 8.485281pt) -- +(45.000000:14.849242pt and 8.485281pt) -- +(135.000000:14.849242pt and 8.485281pt) -- +(225.000000:14.849242pt and 8.485281pt) -- cycle;
\clip (16.500000, -0.000000) +(-45.000000:14.849242pt and 8.485281pt) -- +(45.000000:14.849242pt and 8.485281pt) -- +(135.000000:14.849242pt and 8.485281pt) -- +(225.000000:14.849242pt and 8.485281pt) -- cycle;
\draw (16.500000, -0.000000) node {$\hat{K}_{11}$};
\end{scope}
% Line 5: a G $\hat{D}_{01}$ width=21
\begin{scope}
\draw[fill=white] (49.500000, 15.000000) +(-45.000000:14.849242pt and 8.485281pt) -- +(45.000000:14.849242pt and 8.485281pt) -- +(135.000000:14.849242pt and 8.485281pt) -- +(225.000000:14.849242pt and 8.485281pt) -- cycle;
\clip (49.500000, 15.000000) +(-45.000000:14.849242pt and 8.485281pt) -- +(45.000000:14.849242pt and 8.485281pt) -- +(135.000000:14.849242pt and 8.485281pt) -- +(225.000000:14.849242pt and 8.485281pt) -- cycle;
\draw (49.500000, 15.000000) node {$\hat{D}_{01}$};
\end{scope}
% Line 6: b G $\hat{D}_{11}$ width=21
\begin{scope}
\draw[fill=white] (49.500000, -0.000000) +(-45.000000:14.849242pt and 8.485281pt) -- +(45.000000:14.849242pt and 8.485281pt) -- +(135.000000:14.849242pt and 8.485281pt) -- +(225.000000:14.849242pt and 8.485281pt) -- cycle;
\clip (49.500000, -0.000000) +(-45.000000:14.849242pt and 8.485281pt) -- +(45.000000:14.849242pt and 8.485281pt) -- +(135.000000:14.849242pt and 8.485281pt) -- +(225.000000:14.849242pt and 8.485281pt) -- cycle;
\draw (49.500000, -0.000000) node {$\hat{D}_{11}$};
\end{scope}
% Line 7: a b G $\hat{U}_1$ width=18
\draw (81.000000,15.000000) -- (81.000000,0.000000);
\begin{scope}
\draw[fill=white] (81.000000, 7.500000) +(-45.000000:12.727922pt and 19.091883pt) -- +(45.000000:12.727922pt and 19.091883pt) -- +(135.000000:12.727922pt and 19.091883pt) -- +(225.000000:12.727922pt and 19.091883pt) -- cycle;
\clip (81.000000, 7.500000) +(-45.000000:12.727922pt and 19.091883pt) -- +(45.000000:12.727922pt and 19.091883pt) -- +(135.000000:12.727922pt and 19.091883pt) -- +(225.000000:12.727922pt and 19.091883pt) -- cycle;
\draw (81.000000, 7.500000) node {$\hat{U}_1$};
\end{scope}
% Line 8: a b BARRIER
\draw[decorate,decoration={zigzag,amplitude=1pt,segment length=4}] (96.000000,-7.500000) -- (96.000000,22.500000);
% Line 9: a G $\hat{K}_{0m}$ width=21
\begin{scope}
\draw[fill=white] (112.500000, 15.000000) +(-45.000000:14.849242pt and 8.485281pt) -- +(45.000000:14.849242pt and 8.485281pt) -- +(135.000000:14.849242pt and 8.485281pt) -- +(225.000000:14.849242pt and 8.485281pt) -- cycle;
\clip (112.500000, 15.000000) +(-45.000000:14.849242pt and 8.485281pt) -- +(45.000000:14.849242pt and 8.485281pt) -- +(135.000000:14.849242pt and 8.485281pt) -- +(225.000000:14.849242pt and 8.485281pt) -- cycle;
\draw (112.500000, 15.000000) node {$\hat{K}_{0m}$};
\end{scope}
% Line 10: b G $\hat{K}_{1m}$ width=21
\begin{scope}
\draw[fill=white] (112.500000, -0.000000) +(-45.000000:14.849242pt and 8.485281pt) -- +(45.000000:14.849242pt and 8.485281pt) -- +(135.000000:14.849242pt and 8.485281pt) -- +(225.000000:14.849242pt and 8.485281pt) -- cycle;
\clip (112.500000, -0.000000) +(-45.000000:14.849242pt and 8.485281pt) -- +(45.000000:14.849242pt and 8.485281pt) -- +(135.000000:14.849242pt and 8.485281pt) -- +(225.000000:14.849242pt and 8.485281pt) -- cycle;
\draw (112.500000, -0.000000) node {$\hat{K}_{1m}$};
\end{scope}
% Line 11: a G $\hat{D}_{0m}$ width=21
\begin{scope}
\draw[fill=white] (145.500000, 15.000000) +(-45.000000:14.849242pt and 8.485281pt) -- +(45.000000:14.849242pt and 8.485281pt) -- +(135.000000:14.849242pt and 8.485281pt) -- +(225.000000:14.849242pt and 8.485281pt) -- cycle;
\clip (145.500000, 15.000000) +(-45.000000:14.849242pt and 8.485281pt) -- +(45.000000:14.849242pt and 8.485281pt) -- +(135.000000:14.849242pt and 8.485281pt) -- +(225.000000:14.849242pt and 8.485281pt) -- cycle;
\draw (145.500000, 15.000000) node {$\hat{D}_{0m}$};
\end{scope}
% Line 12: b G $\hat{D}_{1m}$ width=21
\begin{scope}
\draw[fill=white] (145.500000, -0.000000) +(-45.000000:14.849242pt and 8.485281pt) -- +(45.000000:14.849242pt and 8.485281pt) -- +(135.000000:14.849242pt and 8.485281pt) -- +(225.000000:14.849242pt and 8.485281pt) -- cycle;
\clip (145.500000, -0.000000) +(-45.000000:14.849242pt and 8.485281pt) -- +(45.000000:14.849242pt and 8.485281pt) -- +(135.000000:14.849242pt and 8.485281pt) -- +(225.000000:14.849242pt and 8.485281pt) -- cycle;
\draw (145.500000, -0.000000) node {$\hat{D}_{1m}$};
\end{scope}
% Line 13: a b G $\hat{U}_m$ width=18
\draw (177.000000,15.000000) -- (177.000000,0.000000);
\begin{scope}
\draw[fill=white] (177.000000, 7.500000) +(-45.000000:12.727922pt and 19.091883pt) -- +(45.000000:12.727922pt and 19.091883pt) -- +(135.000000:12.727922pt and 19.091883pt) -- +(225.000000:12.727922pt and 19.091883pt) -- cycle;
\clip (177.000000, 7.500000) +(-45.000000:12.727922pt and 19.091883pt) -- +(45.000000:12.727922pt and 19.091883pt) -- +(135.000000:12.727922pt and 19.091883pt) -- +(225.000000:12.727922pt and 19.091883pt) -- cycle;
\draw (177.000000, 7.500000) node {$\hat{U}_m$};
\end{scope}
% Done with gates; drawing ending labels
% Done with ending labels; drawing cut lines and comments
% Done with comments
\end{tikzpicture}
\caption{\label{fig:dku_stages}
The first and final stages of an $m$-stage, two-mode circuit with Kerr and displacement gates equally distributed across both modes, using beamsplitters, $\hat{U}_j$, as entangling operations.
}
\end{figure}
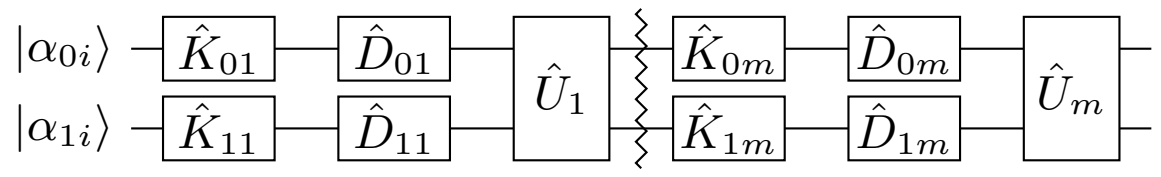

\begin{figure}[t!]
\includegraphics{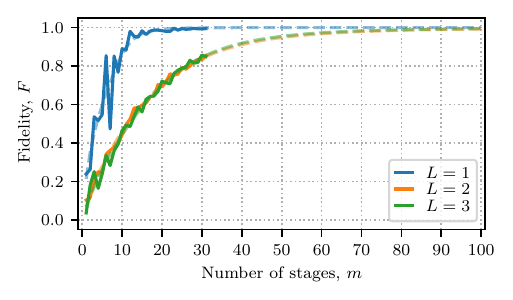}
\caption{\label{fig:bell_opt_fidelities} Solid: fidelities of states to the Bell states $\vert \tilde{\Phi}^+_L \rangle$, generated via the two-mode circuit shown in Fig.~\ref{fig:dku_stages}.
Dashed: fitted curves using a function $g(m) = 1 + 1/f(m)$, where $f(m)$ is a fifth-degree polynomial of positive constants and a degree-$0$ constant of $1$; our choices here were to ensure a monotonic function where $g(0)=0$ and that $g(m)$ is bounded by $1$ for $m \rightarrow \infty$.
In all cases we targeted an amplitude of $\alpha_t=2.5$ to reduce the numerical burden, using fixed Kerr gates $\hat{K}(1/q)$ where $q=2(L+1)$.}
\end{figure}

\begin{figure}[t]
\includegraphics{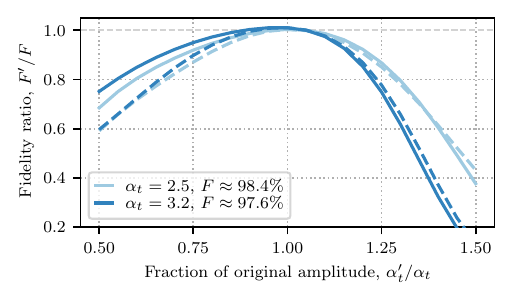}
\caption{\label{fig:bell_param_scaling}
Ratios of fidelities found by scaling a set of parameters for $L=1$ Bell cat generation at two different amplitudes, where $F'$ is the fidelity found via scaling the parameters.
Solid: the initial amplitudes and displacements are scaled according to Eqs.~(\ref{eqn:i_scale}) and (\ref{eqn:d_scale}).
Dashed: only the initial amplitudes are scaled according to Eq.~(\ref{eqn:i_scale}).
}
\end{figure}

To prepare cat-encoded Bell states, we modified the scheme of Fig.~\ref{fig:dk_stages} by introducing beamsplitters at each stage, which function as entangling operations, as shown in Fig.~\ref{fig:dku_stages}.
Because of the increase in parameters per mode and having to simulate two modes in Fock space instead of one, increasing the amplitude leads to a much more significant increase in numerical overhead compared to the single-mode circuit.
Even when holding the Kerr gates fixed to rational parameters, numerical optimisation is very resource intensive, and quickly became a limiting factor in our calculations.

In Fig.~\ref{fig:bell_opt_fidelities} we see the performance for up to $m=31$ stages for the circuit we proposed, using a common target amplitude of $\alpha=2.5$ at all cat code orders, and fixed Kerr gates of $\hat{K}(1/q)$.
For $L=1$ we first exceeded $99.0\%$ fidelity at $m=23$, while for $L>1$ we were unable to achieve high fidelities within our constraints, although the trend with the number of stages suggests that higher fidelities should be possible albeit at a much greater cost in stage count.
Based on the our fitted curves, it appears likely that optimisation would need to be performed on upwards of $100$ stages to achieve near-unit fidelities, which poses problems numerically and physically.

In Fig.~\ref{fig:bell_param_scaling} we show the effects of scaling the involved amplitudes and displacements per the rules outlined in the previous section, albeit considering two different amplitudes.
We had found previously that the scaling effect deteriorates at lower amplitudes, however for the cases we considered here it was only when scaling to lower amplitudes that there was a meaningful benefit, and it remained weaker than we saw for single-mode states.
Considering it is less costly to optimise at a lower amplitude, this provides little practical benefit.
This does not necessarily rule out that some generalisable means of adapting the circuit parameters to new amplitudes, but at least indicates that the intuition from the single-mode circuit cannot be readily ported over.

\section{Conclusion}
\label{sec:conclusion}

We have developed multiple techniques capable of generating high-fidelity many-component cat states, as well as their Bell counterparts.
At the level of single-mode state generation, both our probabilistic and deterministic circuits rely on access to Kerr gates, but show significantly different relationships to how high fidelity can be achieved.
For our Kerr-DFT method, unit fidelity is approached asymptotically in the limit of large amplitude for a fixed, finite circuit, albeit with a probability $q^{q-1}$ for a $q$-component cat state.
For Kerr-displacement, fidelity appears largely independent of amplitude, while unit fidelity is asymptotic in the depth of the circuit, which is complicated by a need for numerical optimisation, although various characteristics make it robust and tunable.

Both of these techniques are theoretically able to produce cat states at realistic sets of parameters, though Kerr-displacement bears a considerable advantage in its ability to achieve high fidelity for a range of different amplitudes.
With respect to circuit depth, Kerr-displacement is also at less of a disadvantage when we consider that Kerr-DFT would need many instances of the same circuit running simultaneously to approach success probabilities meaningfully approaching deterministic.
Gains in either fidelity or savings in circuit depth also may be further possible in Kerr-displacement if more advanced optimisation---perhaps either quasi-analytic or partially in the coherent state basis---is applied, or by inspecting the proximity to fundamental limits \cite{wang2026bosonic}.

For cat-encoded Bell states, our circuit for post-selection perfectly converts two cat states into a corresponding Bell cat state, and is evidently agnostic to how those two resource states are themselves produced.
However it functions only for a subset of the orders of the cat code, and has a rapidly vanishing success probability with no non-zero bound as in our Kerr-DFT scheme.
In the generalisation of the Kerr-displacement scheme, all orders become accessible, but the circuit depth required for high fidelity increases substantially.

Demonstrating the existence and effectiveness of these methods makes a better case for the usefulness of higher-order cat states, on top of the performance gains we had found previously.
Understanding these characteristics of generation helps to contextualise performance gains in terms of overall resource cost.
The most evident bottleneck is in generating cat-encoded Bell states, where either probability or circuit depth become intractable.
Further, current work on the performance of cat-based loss-correction assumes perfect states, and our circuits---for both telecorrection and generation---similarly assume perfect operation, thus both require more robust error models to, in the former case, model the performance of imperfect cat states, and, in the latter, consider imperfect operations.

However, beyond cat states, it is clear that the structural similarity of Kerr states to cat states is what enables much of what we have described, and that inspection of these techniques in the coherent state basis has been essential to much of our understanding.
While for our Kerr-displacement circuit the coherent state basis is too inefficient for direct simulation as the number of stages increases, it still helps to solidify our understanding of unique features, like displacement scaling and codeword switching.
Understanding the evolution of cat states  has benefited from the confluence of various useful characteristics and a favourable basis, and, beyond cats, could inspire the design of codes and techniques designed to take further advantage of them.

\textit{Note added.}
While preparing this manuscript we became aware of related work by H. Kiryu and M. Takeoka \cite{kiryu2026cat}.

\section*{Data Availability}
\label{sec:data}

The results of this paper can be replicated using the Python project \texttt{CoherentStateConstellations} \cite{Boon_CSC_2026}, which includes our numerical implementation of the work presented here, along with a minimal set of Fock space functions originally from \texttt{MrMustard} \cite{mrmustard}.
In said repository we have included scripts to reproduce all data and figures presented herein.

\begin{acknowledgements}
We thank Varun Upreti and Ulysse Chabaud for discussions regarding the Kerr-displacement scheme,
Javier Martínez Cifuentes and Tadeu Tassis for discussions regarding Appendix~\ref{app:pk},
and Filippo M. Miatto for providing early access to a Kerr gate implementation in \texttt{MrMustard}.
\end{acknowledgements}

\section*{Funding}
A.J.B. and N.Q. acknowledge the support from the Ministère de l'Économie et de l'Innovation du Québec, and the Natural Sciences and Engineering Research Council of Canada.
Our research was enabled in part by support provided by the Institut Transdisciplinaire d'Information Quantique (INTRIQ) \doi{10.69777/340940}, a strategic cluster funded by the Fonds de recherche du Québec.

\appendix
\begin{widetext}

\section{Bell cat deformity}
\label{app:bell}

The Bell cat generation protocol we use generalises that of the protocol proposed by Su et al. \cite{su2022universal}, which uses a different basis of cat states which has practical differences from our definition for small amplitudes.
This difference amounts to a deformation in the Bell states in our basis, as in Eq.~\ref{eqn:bell_deformed}.
We will show this using the relationships we derived between these bases in Appendix~B of Ref.~\cite{boon2026generalised}, with which one can write
\begin{equation}
\vert \tilde{J} \rangle_X
=
\frac{1}{\sqrt{2\mathcal{N}}}
\left(
\sqrt{\mathcal{N}_0}
\vert \tilde{0} \rangle_Z
+
(-1)^J
\sqrt{\mathcal{N}_1}
\vert \tilde{1} \rangle_Z
\right),
\end{equation}
where $J=\{0,1\}$, subscripts $X$ denote the choice of basis of Ref.~\cite{su2022universal}, with normalisation constant $\mathcal{N}$, and subscripts $Z$ denote our choice of basis, with two distinct normalisation constants $\mathcal{N}_0$ and $\mathcal{N}_1$.
From here we can write: 
\begin{align}
\vert \tilde{0}, \tilde{0} \rangle_X
+
\vert \tilde{1}, \tilde{1} \rangle_X
&=
\frac{1}{2\mathcal{N}}
\left(
\left[
\mathcal{N}_0
\vert \tilde{0}, \tilde{0} \rangle_Z
+
\sqrt{\mathcal{N}_0 \mathcal{N}_1}
\vert \tilde{0}, \tilde{1} \rangle_Z
+
\sqrt{\mathcal{N}_0 \mathcal{N}_1}
\vert \tilde{1}, \tilde{0} \rangle_Z
+
\mathcal{N}_1
\vert \tilde{1}, \tilde{1} \rangle_Z
\right]
\right)
\notag
\\
&\hphantom{=}+
\frac{1}{2\mathcal{N}}
\left(
\left[
\mathcal{N}_0
\vert \tilde{0}, \tilde{0} \rangle_Z
-
\sqrt{\mathcal{N}_0 \mathcal{N}_1}
\vert \tilde{0}, \tilde{1} \rangle_Z
-
\sqrt{\mathcal{N}_0 \mathcal{N}_1}
\vert \tilde{1}, \tilde{0} \rangle_Z
+
\mathcal{N}_1
\vert \tilde{1}, \tilde{1} \rangle_Z
\right]
\right)
\\
&\propto
\vert \tilde{0}, \tilde{0} \rangle_Z
+
\frac{\mathcal{N}_1}{\mathcal{N}_0}
\vert \tilde{1}, \tilde{1} \rangle_Z.
\end{align}
These normalisation constants can be calculated efficiently via overlaps of coherent states from the definitions of the codewords.

\section{Coefficients of the Kerr decomposition as Gauss sums for $p=\pm1$}
\label{app:sumgauss}

By recognising the coefficients of the Kerr decomposition as Gauss sums, it is possible for us to re-write them explicitly rather than as sums.
We will consider the case $p=-1$ and $q$ even, such that Eq.~(\ref{eqn:gj}) can be simplified to
\begin{equation}
g_{j,-1,q}
=
\frac{1}{q}
\sum^{q-1}_{k=0}
\exp{
\left(
\frac{i \pi k}{q}
\left[
2j
+
k
\right]
\right)}.
\label{eqn:gj1}
\end{equation}
We can complete the square,
\begin{equation}
k(2j+k)
=
k^2 + 2jk
=
(k+j)^2 - j^2,
\end{equation}
to write
\begin{align}
g_{j,-1,q}
&=
e^{-i \pi j^2/q}
\frac{1}{q}
\sum^{q-1}_{k=0}
e^{i \pi (k+j)^2 / q}
\\
&=
e^{-i \pi j^2/q}
\frac{1}{q}
\sum^{q-1}_{k=0}
e^{i \pi k^2 / q}
\label{eqn:gj2}
\end{align}
where in the second line the index $j$ is dropped from the second exponential because it does not affect the sum.
We can identify the sum as a Gauss sum with the following solution for $q$ even \cite{berndt1998gauss},
\begin{equation}
\sum^{q-1}_{k=0}
e^{i \pi k^2 / q} = e^{i \pi /4} \sqrt{q},
\end{equation}
which leaves us with coefficients
\begin{equation}
g_{j,-1,q}
=
e^{-i \pi j^2/q}
\frac{e^{i \pi /4}}{\sqrt{q}}.
\label{eqn:gj3}
\end{equation}
Similarly, for a choice of $p=1$, we find
\begin{equation}
g_{j,1,q}
=
e^{i \pi j^2/q}
\frac{e^{-i \pi /4}}{\sqrt{q}},
\label{eqn:gj3n}
\end{equation}
then for $q$ odd
\begin{equation}
g_{j,\pm 1,q}
=
e^{\pm i \pi (j + 1/2)^2/q}
\frac{e^{\mp i \pi /4}}{\sqrt{q}},
\label{eqn:gj_s_odd}
\end{equation}
so in general we may write
\begin{equation}
g_{j,\pm 1,q}
=
\frac{1}{\sqrt{q}}
{\exp}{\left[
\pm
\frac{i \pi}{q}
\left(
j - \frac{(-1)^q - 1}{4}
\right)^2
\right]}
{\exp}{\left[
\mp
\frac{i \pi}{4}
\right]}.
\label{eqn:gj_s_}
\end{equation}

\section{Conditions for Kerr-DFT cat generation}
\label{app:KDFT}

We will consider the case where $q$ is even and we herald a state in the first mode, meaning our output will be $\vert \tilde{1}_L \rangle$.
The mechanism underlying the Kerr-DFT scheme is the conversion of the input states by the interferometer into a state which contains terms of the form
\begin{equation}
\vert \tilde{1}_L \rangle
\otimes
\vert 0 \rangle^{\otimes (q-1)}
\propto
\left(
\sum^{q-1}_{j=0}
(-1)^j
\vert e^{-2\pi i j /q} \alpha \rangle
\right)
\otimes
\vert 0 \rangle^{\otimes (q-1)}
,
\label{eqn:KDFTout}
\end{equation}
which are the components of the $q$-component cat state in the first mode (written equivalently with a negative sign in the exponential) associated exclusively with vacuum in all other modes, meaning they are heralded exclusively by vacuum measurements on the last $q-1$ modes.
If the other, extraneous terms in the state are associated with non-vacuum coherent states in any of the measured modes, then their contributions will fall off as the amplitudes increase, leading to increasing fidelity to the target $q$-component cat state up to perfect fidelity in the limit of $\alpha \rightarrow \infty$.
Other permutations of Eq.~\ref{eqn:KDFTout} permit heralding of cat states on other modes, though for the sake of brevity we will focus on the case of heralding in the first mode.

By applying the inverse DFT matrix to these terms one can determine what the necessary input states on the DFT interferometer need be to produce this outcome.
With $W^{(q)}$ the DFT matrix of dimension $q$, whose elements are given by
\begin{equation}
W^{(q)}_{j,k}
=
\frac{1}{\sqrt{q}}
e^{-2 \pi i (jk) /q},
\end{equation}
we will define its inverse as $M^{(q)} \equiv \left( W^{(q)} \right)^{-1}$, whose elements are in turn
\begin{equation}
M^{(q)}_{j,k}
=
\frac{1}{\sqrt{q}}
e^{2 \pi i (jk) /q}.
\end{equation}
Applying $\hat{M}^{(q)}$ to Eq.~(\ref{eqn:KDFTout}), we find
\begin{equation}
\hat{M}^{(q)}
\left[
\left(
\sum^{q-1}_{j=0}
(-1)^j
\vert e^{-2\pi i j /q} \alpha \rangle
\right)
\otimes
\vert 0 \rangle^{\otimes (q-1)}
\right]
\propto
\sum^{q-1}_{j=0}
(-1)^j
\vert
e^{-2\pi i j /q} \beta
\rangle^{\otimes q},
\end{equation}
where $\beta = \alpha / \sqrt{q}$.
These terms must occur within our input state for the scheme to work.

Suppose each input mode contains a coherent state $\vert \beta \rangle$ acted on by the Kerr gate $\hat{K}(1/q)$, producing Kerr states of $q$ components,
\begin{equation}
\hat{K}(-1/q) \vert \beta \rangle
=
\sum^{q-1}_{j=0}
g_{j,1,q}
\vert
e^{-2\pi j i /q}
\beta
\rangle,
\label{eqn:Kerr_state}
\end{equation}
which are similar in structure to cat states, in that they are composed of the same coherent states, but do not exhibit the combed Fock support characteristic of cats due to their coefficients $g_j$.

Now consider $q$ modes of these states, which is given by the Cartesian product of the terms in Eq.~(\ref{eqn:Kerr_state}).
The Cartesian product consists of the permutations of all of the terms from each mode, which will necessarily include terms which appear as
\begin{align}
\sum^{q-1}_{j=0}
(g_{j,1,q})^q
\vert
e^{-2 \pi j i /q}
\beta
\rangle^{\otimes q},
\end{align}
where for a given even $q > 0$, all $(g_{j,1,q})^q$ are equal up to an alternating sign, \textit{i.e.},
\begin{align}
\sum^{q-1}_{j=0}
(g_{j,1,q})^q
\vert
e^{-2 \pi j i /q}
\beta
\rangle^{\otimes q}
\propto
\sum^{q-1}_{j=0}
(-1)^j
\vert
e^{-2 \pi j i /q}
\beta
\rangle^{\otimes q},
\end{align}
which is our set of desired input terms.
Since the amplitude of $g_{j,1,q}$ is $1/q$, and these terms contribute $q$ times $|g_{j,1,q}|^q=1/q^q$, this is the origin of the value of the success probability in the limit of large amplitude, $q^{1-q}$.

The remaining terms in the state all contain some coherent light in the non-heralded modes, and thus their contribution to an all-vacuum measurement decreases as amplitude increases, \textit{i.e.}, you are unlikely to measure vacuum from these terms when the coherent states are large.

Additionally, although the conditions to generate states as in Eq.~(\ref{eqn:KDFTout}) are not satisfied by unitaries in general, we have found that the normalised Hadamard matrix using Sylvester's construction \cite{hadamard} can also be used, albeit without the property of being able to switch the codeword generated based on the mode heralded.
While the probability in the limit of large amplitude is tied to the coefficients of the Kerr decomposition, and necessarily independent of the chosen unitary, it is possible that other interferometers may produce other interesting effects on Kerr state inputs.

\section{Probabilistic Bell cats for $L=3$}
\label{app:bell3}

We will consider only the first term in the first mode, $\vert \alpha \rangle$, as the rest generalises to phase.
We want to be left with the correct amplitudes in the second mode, which for $\vert \tilde{\Psi}^{\pm}_3 \rangle$, are given by $\vert e^{\pi i j /4} \alpha \rangle$ for $j=\{1,3,5,7\}$.
When we take the tensor product of our two initial cat states with each other and the vacuum, we have:
\begin{equation}
\begin{matrix}
\hphantom{+}\,\vert& \alpha & 0 & 0 & \alpha & 0 & 0 &\rangle \\
		   +\,\vert& \alpha & 0 & 0 & e^{\pi i/4} \alpha & 0 & 0 &\rangle \\
		   +\,\vert& \alpha & 0 & 0 & e^{\pi i/2} \alpha & 0 & 0 &\rangle \\
		   +\,\vert& \alpha & 0 & 0 & e^{3\pi i/4} \alpha & 0 & 0 &\rangle \\
		   +\,\vert& \alpha & 0 & 0 & -\alpha & 0 & 0 &\rangle \\
		   +\,\vert& \alpha & 0 & 0 & e^{5\pi i/4} \alpha & 0 & 0 &\rangle \\
		   +\,\vert& \alpha & 0 & 0 & e^{3\pi i/2} \alpha & 0 & 0 &\rangle \\
		   +\,\vert& \alpha & 0 & 0 & e^{7\pi i/4} \alpha & 0 & 0 &\rangle 
\end{matrix}.
\label{eqn:bell_i0}
\end{equation}
Then the ${50}{:}{50}$ beamsplitters $\hat{U}$ are applied between mode pairs $(1,2)$, $(1,3)$, $(4,5)$, and $(4,6)$:
\begin{equation}
\begin{matrix}
\hphantom{+}\,\vert& \alpha_2 & \alpha_1 & \alpha_2 & \alpha_2 & \alpha_1 & \alpha_2 &\rangle \\
		   +\,\vert& \alpha_2 & \alpha_1 & \alpha_2 & e^{\pi i/4} \alpha_2 & e^{\pi i/4} \alpha_1 & e^{\pi i/4} \alpha_2 &\rangle \\
		   +\,\vert& \alpha_2 & \alpha_1 & \alpha_2 & e^{\pi i/2} \alpha_2 & e^{\pi i/2} \alpha_1 & e^{\pi i/2} \alpha_2 &\rangle \\
		   +\,\vert& \alpha_2 & \alpha_1 & \alpha_2 & e^{3\pi i/4} \alpha_2 & e^{3\pi i/4} \alpha_1 & e^{3\pi i/4} \alpha_2 &\rangle \\
		   +\,\vert& \alpha_2 & \alpha_1 & \alpha_2 & -\alpha_2 & -\alpha_1 & -\alpha_2 &\rangle \\
		   +\,\vert& \alpha_2 & \alpha_1 & \alpha_2 & e^{5\pi i/4} \alpha_2 & e^{5\pi i/4} \alpha_1 & e^{5\pi i/4} \alpha_2 &\rangle \\
		   +\,\vert& \alpha_2 & \alpha_1 & \alpha_2 & e^{3\pi i/2} \alpha_2 & e^{3\pi i/2} \alpha_1 & e^{3\pi i/2} \alpha_2 &\rangle \\
		   +\,\vert& \alpha_2 & \alpha_1 & \alpha_2 & e^{7\pi i/4} \alpha_2 & e^{7\pi i/4} \alpha_1 & e^{7\pi i/4} \alpha_2 &\rangle 
\end{matrix},
\label{eqn:bell_i1}
\end{equation}
where $\alpha_j \equiv \alpha/ 2^{j/2}$.
Now the rotation $\hat{R}(-\pi/2)$ to the final mode:
\begin{equation}
\begin{matrix}
\hphantom{+}\,\vert& \alpha_2 & \alpha_1 & \alpha_2 & 			   \alpha_2 &   		   \alpha_1 & e^{3\pi i/2} \alpha_2 &\rangle \\
		   +\,\vert& \alpha_2 & \alpha_1 & \alpha_2 &  e^{\pi i/4} \alpha_2 &  e^{\pi i/4} \alpha_1 & e^{7\pi i/4} \alpha_2 &\rangle \\
		   +\,\vert& \alpha_2 & \alpha_1 & \alpha_2 &  e^{\pi i/2} \alpha_2 &  e^{\pi i/2} \alpha_1 &  \alpha_2 &\rangle \\
		   +\,\vert& \alpha_2 & \alpha_1 & \alpha_2 & e^{3\pi i/4} \alpha_2 & e^{3\pi i/4} \alpha_1 & e^{\pi i/4} \alpha_2 &\rangle \\
		   +\,\vert& \alpha_2 & \alpha_1 & \alpha_2 &    		  -\alpha_2 & 			  -\alpha_1 & e^{\pi i/2} \alpha_2 &\rangle \\
		   +\,\vert& \alpha_2 & \alpha_1 & \alpha_2 & e^{5\pi i/4} \alpha_2 & e^{5\pi i/4} \alpha_1 & e^{3\pi i/4} \alpha_2 &\rangle \\
		   +\,\vert& \alpha_2 & \alpha_1 & \alpha_2 & e^{3\pi i/2} \alpha_2 & e^{3\pi i/2} \alpha_1 &             -\alpha_2 &\rangle \\
		   +\,\vert& \alpha_2 & \alpha_1 & \alpha_2 & e^{7\pi i/4} \alpha_2 & e^{7\pi i/4} \alpha_1 & e^{5\pi i/4} \alpha_2 &\rangle 
\end{matrix},
\label{eqn:bell_i2}
\end{equation}
and from here it should be apparent that wherever there are pairs of equal or opposite amplitudes, like $(\beta,\beta)$ or $(\beta,-\beta)$, between modes $2$ and $5$ and modes $3$ and $6$, there will be cancellation from the application of a ${50}{:}{50}$ beamsplitter to those modes, where
\begin{equation}
\hat{U} \vert \beta, \beta \rangle
=
\biggl\vert \sqrt{2} \beta, 0 \biggr\rangle,
\quad
\hat{U} \vert \beta, -\beta \rangle
=
\biggl\vert 0, \sqrt{2} \beta \biggr\rangle.
\end{equation}
This will lead to vacuum in those modes, and thus the condition that if vacuum is not measured the associated terms will not appear in the output.

We see here that after measurement only the first and third modes will remain, with the remaining terms being the second, fourth, sixth, and eighth from Eq.~(\ref{eqn:bell_i2}), which correspond to our desired terms listed previously.
The condition requiring that all photon-number measurements need be equal arise from the fact that we require the coefficients---which determined by our measurements---to be equal across all of the terms in order for the state to be proportional to a Bell state.

\section{Displacement sign-flips for codeword switching in displacement-Kerr cat generation}
\label{app:pk}

In Section~\ref{sec:inline_qcats} we noted that flipping the signs for all odd or even displacements would, for $\alpha_i \gg \beta_j$, take us from a state with some fidelity to $\vert \tilde{0}_L \rangle$ to a state with an approximately equal fidelity to $\vert \tilde{1}_L \rangle$.
We may use the result from Appendix~\ref{app:sumgauss} to inspect how the parity operator interacts with Kerr gates of even $q$.
While here we will explicitly consider $\hat{K}(1/q)$, its relationship to $\hat{K}(-1/q)$ is merely $\hat{R}(\pi/q) \hat{K}(1/q) = \hat{K}(-1/q)$, where $\hat{R}(\pi/q)$ is the Pauli $Z$ for cat codes of even $q$, and so the mechanism is fundamentally unchanged.

The sign flips in the displacements can be written in terms of the parity operator,
\begin{equation}
\hat{\Pi} \equiv \hat{R}(\pi/2)
\end{equation}
where for either all odd or even $j$ we have
\begin{equation}
\hat{\Pi} \hat{D}(\beta_j) \hat{\Pi}
=
\hat{D}(-\beta_j).
\end{equation}
Given the scheme from Fig.~\ref{fig:dk_stages}, for sign flips on every odd $j^{\text{th}}$ displacement up to a total number $m$ odd, we can write a series of operations in the circuit as
\begin{equation}
\left[ \hat{\Pi} \hat{D}_m \hat{\Pi} \right]
\hat{K}
\hat{D}_{m-1}
\hat{K}
\left[ \hat{\Pi} \hat{D}_{m-2} \hat{\Pi} \right]
\hat{K}
\cdots
\hat{D}_{2}
\hat{K}
\left[ \hat{\Pi} \hat{D}_{1} \hat{\Pi} \right]
\hat{K}
\hat{D}_0,
\end{equation}
where $\hat{D}_j \equiv \hat{D}(\beta_j)$ and  $\hat{K} \equiv \hat{K}(1/q)$, and an equivalent expression can be written for even flips up to an overall parity operator.
Now instead of applying the parity operator to the displacements, we can instead pair each one with a Kerr gate (up to an overall parity) and define these modified Kerr gates as
\begin{equation}
\hat{K}_{\Pi}(1/q) \equiv \hat{\Pi} \hat{K}(1/q) = \hat{K}(1/q) \hat{\Pi},
\end{equation}
where we have used the fact that $\hat{\Pi}$ and $\hat{K}$ commute.
Applying the parity operator to $\hat{K}(1/q)$ for even $q$ can be understood as a shifting of the indices of the coefficients (leaving the indices besides $j$ implicit):
\begin{align}
\hat{\Pi} \hat{K}(1/q)
&=
e^{i \pi \hat{n}}
\sum^{q-1}_{j=0}
g_j
e^{- i j \pi \hat{n} / q}
=
\sum^{q-1}_{j=0}
g_j
e^{- i (j - q/2) \pi \hat{n} / q}
=
\sum^{q-1}_{j=0}
g_{j + q/2}
e^{- i j \pi \hat{n} / q}.
\end{align}
Using Eq.~(\ref{eqn:gj3}), we can write
\begin{equation}
g_{j+q/2}
=
e^{-i \pi (j + q/2)^2/q}
\frac{e^{i \pi /4}}{\sqrt{q}},
\label{eqn:gj3_}
\end{equation}
from which we find a change to the coefficients of even and odd $j$ shown in Table~\ref{tab:indices}.

\begin{table}[t]
\caption{\label{tab:indices}
The factors picked up in the the transformation $\hat{K}(1/q) \rightarrow \hat{K}_{\Pi}(1/q)$, which takes the Kerr decomposition coefficients $g_j \rightarrow c g_{j+q/2}$ for even $q$, and $c$ some scalar coefficient.}
\centering
\begin{tabular}{c | r | r | r}
\hline
$q \pmod{8}$ & $j$ even & $j$ odd & overall $c$ \\
\hline
$0$ & $ 1$ & $-1$ & $ 1$ \\
$2$ & $ i$ & $-i$ & $ i$ \\
$4$ & $-1$ & $ 1$ & $-1$ \\
$6$ & $-i$ & $ i$ & $-i$ \\
\end{tabular}
\end{table}

From Table~\ref{tab:indices} we can identify three operations at play that relate to our logical basis $\vert \tilde{J} \rangle$: the phase gate $S$, the Pauli $Z$, and a change of sign $1 \rightarrow (-1)^j$.
The phase gate is responsible for an overall $i$ for $q \pmod{8} = \{2,6\}$, while Pauli $Z$ applies an overall sign change in $q \pmod{8} = \{4,6\}$; these two operations have no impact on our fidelity to the Pauli $Z$ eigenstates.

It is $1 \rightarrow (-1)^j$, observed for all $q$, which is insightful, as this is the relationship between the coefficients of $\vert \tilde{0} \rangle$ and $\vert \tilde{1} \rangle$, thus the relationship between the operators $\hat{K}(1/q)$ and $\hat{K}_{\Pi}(1/q)$ can be thought of as like that between the cat code's Pauli $Z$ eigenstates.
This relationship provides an alternative intuition for what we observe about displacement flips, in that if a series of operations $\hat{K}(1/q)$ and $\hat{D}(\beta_j)$ produces a $\vert \tilde{0} \rangle$-like state, then $\hat{K}_{\Pi}(1/q)$ and $\hat{D}(\beta_j)$ will produce a state that is $\vert \tilde{1} \rangle$-like.

From here we can inspect the overlap more specifically, where our argument is as follows:
we can consider the only significantly contributing overlaps to be between terms which overlap exactly up to the approximation of the displacements being sufficiently small relative to the amplitudes of the initial coherent states;
under this approximation, considering these terms alone, we find that in the overlap with $\vert \tilde{1}_L \rangle$, the terms from the cat state with a factor $(-1)$ correspond to terms with factor $(-1)$ in the state produced by the Kerr-displacement protocol with odd displacements flipped in sign, leading to cancellation of these factors;
as these terms are otherwise identical to those of $\vert \tilde{0}_L \rangle$ and the Kerr-displacement states sans sign flips, the total overlaps are therefore identical.
We will now follow this argument in detail.

To avoid the trouble of the growth in our number of terms from the Kerr-displacement protocol, which grows as $q^m$ for $m$ stages, let us only consider a single term decided by a subset of the rotations from the decomposed Kerr gates, where we will denote each rotation as
\begin{equation}
\hat{R}(j_n \phi_q),
\end{equation}
for $j_n$ the selected index at a stage $n$, with $\phi_q \equiv 2 \pi / q$, and $q=2(L+1)$.
Overall, the influence of the parity operators on these rotations will be an overall sign on the final term, determined by
\begin{equation}
s=
(-1)^{\sum^{m}_{n=1} j_n},
\label{eqn:overall_sign}
\end{equation}
where for the sum of the indices $\sum^{m}_{n=1} j_n$ even, the additional sign on the term is $(+1)$, while for the sum odd it is $(-1)$.
Any additional operations, as seen in Table~\ref{tab:indices}, will amount only to constants times the entire state, and thus can be ignored for our purposes.

Using our subset of rotations, we may write the term associated with this set of rotations, where we begin with our coherent state as usual and apply the rotations interleaved with displacements,
\begin{align}
\left(
\prod^{m}_{n=1}
\hat{D}(\beta_n)
\hat{R}(j_n \phi_q)
\right)
\vert \beta_0 \rangle
=
s
e^{i \phi}
\left\vert
\prod^{m}_{n=1}
e^{i j_n \phi_q}
\beta_0
+
\beta_m
+
\sum^{m-1}_{k=1}
\left[
\prod^{k}_{n=1}
e^{i j_n \phi_q}
\right]
\beta_{m-n}
\right\rangle
\approx
s
e^{i \phi}
\vert
r \alpha
+
B
\rangle,
\end{align}
where our only approximation is in taking $\beta_0 \approx \alpha \in \mathbb{R}$ for simplicity, and otherwise $e^{i \phi}$ is a phase factor picked up from each displacement, while we have written the sum of the rotated displacements as
\begin{equation}
B
\equiv
\beta_m
+
\sum^{m-1}_{k=1}
\left[
\prod^{k}_{n=1}
e^{i j_n \phi_q}
\right]
\beta_{m-n},
\label{eqn:rot_term}
\end{equation}
and the product of all rotations as
\begin{equation}
r
\equiv
\prod^{m}_{n=1}
e^{i j_n \phi_q}.
\end{equation}
The phase factor $e^{i \phi}$ depends only on the displacements and is independent of our choice of $\hat{K}$ or $\hat{K}_{\Pi}$, whose choice only determines the overall factor $s$, thus going forward $e^{i \phi}$ may be ignored in our comparison.
Moreover, there is also a product of coefficients from the Kerr decomposition associated with this term, but it is determined solely by the choice of $j_n$, and again independent of $\hat{K}$ or $\hat{K}_{\Pi}$, allowing us to ignore it in our comparison of the circuit between the choice of $\hat{K}$ or $\hat{K}_{\Pi}$.

Let us first consider our $\alpha$ relative to $B$, specifically the overlap of a coherent state to itself plus the displacements $B$, using Eq.~\ref{eqn:coherent_overlap} and $\alpha \gg B$,
\begin{align}
\langle \alpha \vert \alpha + B \rangle
&=
{\exp}{
\left[
\frac{1}{2}
\left(
\alpha
B
-
\alpha
B^*
-
\vert
B
\vert^2
\right)
\right]
}
\approx
{\exp}{
\left[0\right]}
=
1
=
\langle \alpha \vert \alpha \rangle
\end{align}
where, in our case, the maximum possible value of any $B$ is given by
\begin{equation}
B_{\text{max}} = \sum^{m}_{j=1} | \beta_j |^2.
\end{equation}
Now consider the overlap between a coherent state and a rotated counterpart at phase $x \phi_q = 2 \pi x / q$ for $x \in \mathbb{Z}$,
\begin{align}
\langle e^{i x \phi_q} \alpha \vert \alpha \rangle
&=
{\exp}{\left[ \frac{1}{2} \left( e^{-i x \phi_q} \alpha^2 - e^{i x \phi_q} \alpha^2 - \alpha^2 \vert e^{i x \phi_q} - 1 \vert^2 \right) \right]}
\\
&=
{\exp}{\left[ \frac{1}{2} \left( e^{-i x \phi_q} \alpha^2 - e^{i x \phi_q} \alpha^2 - \alpha^2 (e^{i x \phi_q} - 1)(e^{-i x \phi_q} - 1) \right) \right]}
\\
&=
{\exp}{\left[ e^{-i x \phi_q} \alpha^2 \right]},
\end{align}
where, in order to ignore terms that are not exactly overlapping, we want
\begin{equation}
\vert \langle \alpha \vert \alpha \rangle \vert^2
=
1
\gg
\left| \langle e^{2 i \pi / q} \alpha \vert \alpha \rangle \right|^2
=
\left| {\exp}{\left[ e^{-2 i \pi / q} \alpha^2 \right]} \right|^2
=
{\exp}{\left[ 2 \alpha^2 \left( {\cos}{(2 \pi / q)} - 1 \right) \right]},
\label{eqn:overlap_q}
\end{equation}
where we see that since ${\cos}{(2 \pi / q)} - 1 < 0$ we can always choose sufficiently large $\alpha$ for the exponential to be small.
With this we then know that for sufficiently large $\alpha$, the only significant overlaps $\langle e^{i x \phi_q} \alpha \vert \alpha \rangle$ will be where $x=0$, \textit{i.e.}, where the phase of the bra matches that of the ket.

Now we take the overlap of Eq.~(\ref{eqn:rot_term}) to a term of a $q$-component cat state, taking the $x^{\text{th}}$ component, which can be written as
\begin{align}
\langle
e^{i x \phi_q} \alpha
\vert
r \alpha + B
\rangle
&=
{\exp}{
\left[
\frac{1}{2}
\left(
e^{-i x \phi_q}
r
\alpha^2
-
e^{i x \phi_q}
r^*
\alpha^2
+
e^{-i x \phi_q}
\alpha
B
-
e^{i x \phi_q}
\alpha
B^*
-
\vert
e^{i x \phi_q} \alpha - r \alpha - B
\vert^2
\right)
\right]
}.
\end{align}
Consider that for sufficiently large $\alpha$, and given our set of rotations $q$, per Eq.~(\ref{eqn:overlap_q}), we will only have significant overlaps between $r \alpha$ and $e^{i x \phi_q} \alpha$ when $r = e^{i x \phi_q}$, \textit{i.e.},
\begin{equation}
\left[ \sum^{m}_{n=1} j_n \right]
=
x \pmod{q}.
\label{eqn:rotation_condition}
\end{equation}
Thus if we only consider these strongly-overlapping sets of rotations, we can write
\begin{align}
\langle
e^{i x \phi_q} \alpha
\vert
r \alpha + B
\rangle
&\approx
{\exp}{
\left[
\frac{1}{2}
\left(
e^{-i x \phi_q}
\alpha
B
-
e^{i x \phi_q}
\alpha
B^*
\right)
\right]
},
\label{eqn:approx_overlap}
\end{align}
where, beyond outright cancellations, we have taken $| B |^2$ to be negligible using our assumption that the maximum possible total displacement $B$ is small relative to $\alpha$.

Now, Eq.~(\ref{eqn:approx_overlap}) describes our overlap to a term in an even-$q$ cat state irrespective of which Pauli $Z$ eigenstate it is, and relies only on the condition of Eq.~(\ref{eqn:rotation_condition}).
What we can take from Eq.~(\ref{eqn:rotation_condition}) is that $x$ and the sum over the indices $j_n$ necessarily have equal parity, \textit{i.e.}, they must be simultaneously either even or odd.
Thus
\begin{equation}
\left[
x + \sum^{m}_{n=1} j_n
\right]
= 0 \pmod{2},
\end{equation}
\textit{i.e.}, said sum is necessarily even.
Knowing what we do about the sign of our Kerr-displacement terms from Eq.~(\ref{eqn:overall_sign}), and considering that the cat terms have a sign $(-1)^x$ for $\vert e^{i x \phi_q} \alpha \rangle$, then under these conditions we have a constant on each term of the complete overlap given by
\begin{equation}
s (-1)^x
=
(-1)^{x + \sum^{m}_{n=1} j_n}
=
1.
\end{equation}
This means that every term of the overlap, given by Eq.~(\ref{eqn:approx_overlap}), remain approximately unchanged in the approximation of small total displacement (up to an overall constant) between either choice of Kerr gates $\hat{K}(1/q)$ and overlap to $\vert \tilde{0}_L \rangle$, or Kerr gates $\hat{K}_{\Pi}(1/q)$ and overlap to $\vert \tilde{1}_L \rangle$,
\begin{equation}
\left\vert
\langle
\tilde{0}_L
\vert
\left(
\prod^{m}_{j=1}
\hat{D}(\beta_j)
\hat{K}(1/q)
\right)
\vert
\alpha
\rangle
\right\vert^2
\approx
\left\vert
\langle
\tilde{1}_L
\vert
\left(
\prod^{m}_{j=1}
\hat{D}(\beta_j)
\hat{K}_{\Pi}(1/q)
\right)
\vert
\alpha
\rangle
\right\vert^2.
\end{equation}
Moreover, we know that swapping $\hat{K}$ for $\hat{K}_{\Pi}$ is identical to sign flips on every odd displacement,
\begin{equation}
\left\vert
\langle
\tilde{1}_L
\vert
\left(
\prod^{m}_{j=1}
\hat{D}(\beta_j)
\hat{K}_{\Pi}(1/q)
\right)
\vert
\alpha
\rangle
\right\vert^2
=
\left\vert
\langle
\tilde{1}_L
\vert
\left(
\prod^{m}_{j=1}
\hat{D}((-1)^j \beta_j)
\hat{K}(1/q)
\right)
\vert
\alpha
\rangle
\right\vert^2,
\end{equation}
and so we may assert that
\begin{equation}
\left\vert
\langle
\tilde{0}_L
\vert
\left(
\prod^{m}_{j=1}
\hat{D}(\beta_j)
\hat{K}(1/q)
\right)
\vert
\alpha
\rangle
\right\vert^2
\approx
\left\vert
\langle
\tilde{1}_L
\vert
\left(
\prod^{m}_{j=1}
\hat{D}((-1)^j \beta_j)
\hat{K}(1/q)
\right)
\vert
\alpha
\rangle
\right\vert^2,
\end{equation}
\textit{i.e.}, that flipping the signs on the odd displacements in the Kerr-displacement circuit produces a state of approximately equal fidelity to $\vert \tilde{1}_L \rangle$ as it does without sign flips to $\vert \tilde{0}_L \rangle$.

\end{widetext}

\bibliography{refs}

\end{document}